\documentclass[fleqn,usenatbib]{mnras}

\usepackage{tikz}
\def\checkmark{\tikz\fill[scale=0.4](0,.35) -- (.25,0) -- (1,.7) -- (.25,.15) -- cycle;} 
\usepackage{graphicx}

\usepackage[utf8x]{inputenc}
\usepackage{color}
\usepackage{txfonts}
\usepackage{natbib}
\usepackage[squaren,Gray]{SIunits}

\usepackage{amsmath}

\newcommand{\gcc}{\mbox{ g cm$^{-3}$}}
\def\nh{n_\mathrm{H}}
  \renewcommand{\vec}[1]{\mathbf{#1}}

\usepackage{bbold}
\usepackage{subcaption}
\usepackage{mwe}

\title[Dust evolution during the protostellar collapse.]{Protostellar collapse simulations in spherical geometry with dust coagulation and fragmentation.}

\author[Lebreuilly et al.]{
Ugo Lebreuilly,$^{1}$\thanks{E-mail: ugo.lebreuilly@cea.fr}
Valentin Vallucci-Goy,$^{1}$
       Vincent Guillet, $^{2,3}$
          Maxime Lombart,$^{4}$
      and  Pierre Marchand $^{5}$
\\
$^{1}$Universit\'{e} Paris-Saclay, Universit\'{e} Paris Cité, CEA, CNRS, AIM, 91191, Gif-sur-Yvette, France\\
$^{2}$  Institut d’Astrophysique Spatiale, CNRS, Universit\'{e} Paris-Saclay, CNRS, B\^{a}t. 121, 91405 Orsay, France \\
$^{3}$ Laboratoire Univers et Particules de Montpellier, Universit\'{e} de Montpellier, CNRS/IN2P3, CC 72, Place Eug\`{e}ne Bataillon, 34095 Montpellier Cedex 5, France\\
$^{4}$  Department of Earth Sciences, National Taiwan Normal University, 88, Sec.4, Ting-Chou Road, Taipei 11677, Taiwan \\
$^{5}$  Institut de Recherche en Astrophysique et Plan\'{e}tologie, Universit\'{e} Paul Sabatier Toulouse 3, 118 Rte de Narbonne, 31062 Toulouse, France
}
\date{Accepted XXX. Received YYY; in original form ZZZ}

\pubyear{2022}

\begin{document}
\label{firstpage}
\pagerange{\pageref{firstpage}--\pageref{lastpage}}
\maketitle
\begin{abstract}
  We model the coagulation and fragmentation of dust grains during the protostellar collapse with our newly developed \texttt{shark} code. It solves the gas-dust hydrodynamics in a spherical geometry and the coagulation/fragmentation equation. It also computes the ionization state of the cloud and the Ohmic, ambipolar and Hall resistivities. 
   We find that the dust size distribution evolves significantly during the collapse, large grain formation being controlled by the turbulent differential velocity. When turbulence is included, only ambipolar diffusion remains efficient at removing the small grains from the distribution, brownian motion is only efficient as a standalone process. The macroscopic gas-dust drift is negligible for grain growth and only dynamically significant near the first Larson core. At high density, we find that the coagulated distribution is unaffected by the initial choice of dust distribution. Strong magnetic fields are found to enhance the small grains depletion, causing an important increase of the ambipolar diffusion. This hints that the magnetic field strength could be regulated by the small grain population during the protostellar collapse. Fragmentation could be effective for bare silicates, but its modeling relies on the choice of ill-constrained parameters. It is also found to be negligible for icy grains. When fragmentation occurs, it strongly affects the magnetic resistivities profiles.  
   Dust coagulation is a critical process that needs to be fully taken into account during the protostellar collapse. The onset and feedback of fragmentation remains uncertain and its modeling should be further investigated.
\end {abstract}

\section{Introduction}

Despite representing only about $\sim 1 \%$ of the mass of the interstellar medium, dust grains are among its essential components. They play a fundamental role in the cooling of star forming clouds through their absorption and thermal emission \citep{McKeeOstriker2007}, as well as their chemistry as privileged formation site of H$_2$ \citep{GouldSalpeter1963}. In addition, dust grains have a major impact on the coupling between the neutrals and the magnetic field through the magnetic resistivities \citep[see for e.g.,][]{Marchand2016,Zhao2016,Marchand2021,Tuskamoto2022}. Last but not least, the grains are the building blocks of planets that are expected to be formed by the coagulation and local accumulation of dust in protoplanetary disks \citep[see the review by][]{Testi2014}.

The dust size distribution is usually modelled as a Mathis, Rumpl, Nordsieck (MRN) distribution \citep{MRN}, a power law distribution that ranges between a few nanometers and less than a micron designed to reproduce the dust component of the diffuse ISM. However, in addition from being debated in the diffuse ISM itself \citep{Jones2013,Kohler2015,Jones2017}, the MRN is most likely incorrect in the denser regions, e.g. in molecular clouds, prestellar cores and protoplanetary disks. Recent observations indeed seem to indicate that the dust grains are growing significantly prior to the protoplanetary disk phase in the ISM \citep[see for e.g.][]{Pagani2010,Kataoka2015,Galametz2019,Valdivia19}. The theoretical works \citep{Ormel2009,HirashitaOkumai2009,Vorobyov2019,Silsbee2020,Guillet2020,Marchand2021,Tsukamoto2021,Kawasaki2022,Bate2022,Tu2022} that take into account the dust grain growth (and sometimes fragmentation) all reach the same conclusion: dust grains are growing in collapsing protostellar cores. 

A long standing problem of the disk formation is the so-called magnetic braking catastrophe. In the ideal MHD limit, the magnetic braking is so strong during the protostellar collapse that it could prevent the formation of a disk supported by its rotation. The inclusion of non-ideal MHD effects is a promising solution for this problem \citep[see for e.g.,][]{Li2014,Masson2016,Machida2016,Wurster2016,Hennebelle2020}. However, recent studies have shown that the outcome of the collapse of a rotating core could significantly depend on the choice of dust distribution through its impact on the resistivities. \cite{Zhao2016} have shown, for example, that the removal of the smallest grains could promote the formation of a supported disk. Similar findings have been reported by \cite{Marchand2020}, where they also note an impact of the dust size distribution on the outflow and on the fragmentation of the cores.

Recently, dust evolution during the collapse of protostellar cores has been studied either with single-zone collapse models or through multidimensional simulations. Simulations can follow grain-grain coagulation, either solving the Smoluchowski \citep{Smolu16} equation \citep{Vorobyov2019,Bate2022,Tu2022} or using a monodisperse model \citep{Tsukamoto2021}. The simplicity of single-zone models allow for the inclusion of more complex dust physics, such as the feedback of dust evolution on the evolution of magnetic resistivities during the collapse, with detailed MHD grain dynamics \citep{Silsbee2020,Guillet2020} and recently with grain fragmentation \citep{Kawasaki2022}.

In this work, we aim to make a step forward toward more accurate simulations that self-consistently account for both dust evolution and non-ideal MHD effects. We therefore present new hydrodynamical simulations for the spherically symmetric collapse of protostellar clouds that include gas and multiple grain species and account for the competition between coagulation and fragmentation of dust grains with an "on-the-fly" calculation of the resistivities as per \cite{Marchand2021}. We compare the relative importance of the initial dust size distribution, of the initial magnetic field strength, and of the various mechanisms responsible for grain dynamics and evolution, in determining the evolution of the dust size distribution and magnetic properties of the core through the collapse. We made a particular effort in the modeling of fragmentation with the goal of delimiting the relatively uncertain consequences of this collisional process on the dust size distribution and thereby the evolution of magnetic resistivities.

This paper is arranged as followed. In Sect.~\ref{sec:context}, we recall the theoretical context of our work. Then, in Sect.~\ref{sec:method}, we introduce the \texttt{shark} code and the numerical methods used in this work. Our simulation results are then described in Sect.~\ref{sec:results} and discussed in Sect.~\ref{sec:discussion}. Finally, we present our conclusions in  Sect.~\ref{sec:conclusion}.
\section{Physical model}
\label{sec:context}

\subsection{Gas and dust hydrodynamical equations}
\label{sec:HydroEq}
Let us consider a gas and dust mixture with $\mathcal{N}$ different grain sizes. In the context of the protostellar collapse, neglecting the impact of magnetic fields on the gas and dust motions, the equation of gas and dust hydrodynamics can be written as

\begin{align}
 \frac{\partial \rho}{\partial t}  + \nabla \cdot \left[ \rho \vec{v} \right]&=&0, \nonumber  \\   \frac{\partial   \rho_{k} }{\partial t} +\nabla \cdot \left[ \rho_{k} \vec{v_k} \right] &=& S_{k,\mathrm{growth}}, \ \forall k \in \left[1,\mathcal{N}\right], \nonumber \\
  \frac{\partial \rho \vec{v}}{\partial t}  + \nabla \cdot \left[ \rho \vec{v} \vec{v} + P \mathbb{I} \right]&=& \rho \vec{g} + \sum_j \
  \frac{\rho_j}{t_{\rm{s},j}}  (\vec{v}_j-\vec{v}),\nonumber \\
  \frac{\partial \rho_k \vec{v_k}}{\partial t}  + \nabla \cdot \left[ \rho_k \vec{v}_k \vec{v}_k  \right]&=& \rho_k \vec{g}  - \frac{\rho_k}{t_{\rm{s},k}} (\vec{v}_k-\vec{v}),
  \label{eq:hydro}
\end{align}
where $\rho$ and $\vec{v}$ are the gas density and velocity, $P$ the gas thermal pressure, and $\vec{g}$ the gravitational acceleration. Regarding dust properties, $\rho_k$, $\vec{v_K}$ and $t_{\rm{s},k}$, are the dust mass density, grain velocity and grain stopping time \citep{Epstein1924} for the dust species $k$, respectively, while $S_{k,\mathrm{growth}}$ is the source terms due to the coagulation and fragmentation of dust grains in grain-grain collisions (see Section \ref{sec:evolution}).

\subsection{Dust differential velocities}
\label{sec:drifts}
Let us now summarize the four different sources of dust differential velocities considered in our work.
\subsubsection{Turbulent differential velocity}

The gas turbulence cascade can accelerate dust grains \citep{Voelk1980}. The efficiency of this process depends both on the characteristic of the turbulence (such as its Reynolds number $\mathrm{Re}$) and on the grain size.
To model this mechanism, we follow \cite{Guillet2020} and use the frame defined by \cite{Ormel2009} as well as the expressions for the grain turbulent differential velocity derived by \cite{Ormel2007}.

The timescale of injection of the turbulence is considered to be the free-fall timescale
\begin{equation}
t_{\mathrm{ff}} =\sqrt{\frac{3\pi}{32 \mathcal{G}\rho}}.
\end{equation}
The timescale of the dissipation of turbulence $t_{\eta}$ is by definition 
\begin{equation}
t_{\eta}= t_{\mathrm{ff}}/\sqrt{\mathrm{Re}}\,,
\end{equation}
 where \begin{equation}
\mathrm{Re} = 6.2 \times 10^{7} \sqrt{\frac{n_{\mathrm{H}}}{10^{5}~\centi\meter^{-3}}} \sqrt{\frac{T }{10~K}}.
\label{eq:Re}
\end{equation}
as per \cite{Ormel2009}.

For a grain size $i$ of radius $s_i$, knowing the gas sound speed $c_{\mathrm{s}} \equiv \sqrt{\gamma P/\rho}$, we derive the grain stopping time \citep{Epstein1924}
\begin{equation}
t_{\rm{s},i} \equiv \sqrt{\frac{\pi \gamma}{8}} \frac{\rho_{\rm{grain}} s_i}{\rho \mathrm{c_{\rm{s}}}}\,,
\end{equation}
from which we define the grain Stokes number $\mathrm{St}_i\equiv t_{\rm{s},i}/t_{\mathrm{ff}}$.

The relative velocity $\Delta  v_{\mathrm{turb,i,j}}$ induced by turbulence between grains $i$ and $j$ is responsible for their collisions. From now on, and for the remaining of the paper, grain $i$ will be the larger grain and grain $j$ the smaller grain. Depending on the grains stopping time, the interaction between turbulence and grains is described by three types of vortices (class I, II or III vortices), leading to three different expressions for $\Delta  v_{\mathrm{turb,i,j}}$
\begin{equation}
\Delta v_{\mathrm{turb,i,j}}^2 \equiv \left\{
    \begin{array}{ll}
       V_{\mathrm{g}}^2 \, \frac{\mathrm{St}_i-\mathrm{St}_j}{\mathrm{St}_i+\mathrm{St}_j} (\frac{\mathrm{St}_i^2}{\mathrm{St}_i+1/\sqrt{\mathrm{Re}}}+\frac{\mathrm{St}_j^2}{\mathrm{St}_j+1/\sqrt{\mathrm{Re}}}), \mbox{~if~} t_{\rm{s},i} < t_{\eta} \\
        V_{\mathrm{g}}^2  \, \beta_{\mathrm{i,j}} \mathrm{St}_i   \mbox{,~if~}  t_{\eta} \le t_{\rm{s},i} < t_{\mathrm{ff}} \\
      V_{\mathrm{g}}^2  \, (\frac{1}{\mathrm{St}_i+1}+\frac{1}{\mathrm{St}_j+1}) \mbox{, otherwise,}
        \end{array}
        \right.
\end{equation}
where $ V_{\mathrm{g}} = \sqrt{3/2}c_{\rm{s}}$ \citep{Guillet2020,Kawasaki2022} and $\beta_{\mathrm{i,j}}=3.2 -(1+x_{i,j})+\frac{2}{1+x_{i,j}}(\frac{1}{2.6}+\frac{x_{i,j}^3}{1.6+x_{i,j}})$, $x_{i,j}$ being the ratio between $\mathrm{St}_i$ and $\mathrm{St}_j$.

The intermediate regime ($t_{\eta} \le t_{\rm{s},i} < t_{\mathrm{ff}}$), which is valid for a large range of grain sizes in our study, is known to be quite efficient at growing large grains. Depending on the Reynolds number, turbulence could be the main source of grain growth during the protostellar collapse \citep{Silsbee2020,Guillet2020}. We note that the framework of \cite{Ormel2007}, valid for a Kolmogorov cascade, has been recently generalised to arbitrary turbulent cascades by \cite{Gong2020,Gong2021}.

\subsubsection{Hydrodynamical drift}
The drift between gas and dust, which is size-dependent, is responsible for a relative velocity between grains sizes $i$ and $j$ that we call
the hydrodynamical drift velocity $\Delta  v_{\mathrm{hydro},i,j}$:
\begin{equation}
\Delta  v_{\mathrm{hydro},i,j} \equiv |\vec{v}_i-\vec{v}_j|.
\end{equation}
 This drift is self-consistently determined  by {\ttfamily shark} by solving the gas and dust dynamical equations presented in Sect. \ref{sec:HydroEq}.

 The terminal velocity approximation is most probably valid during the protostellar collapse \citep{Lebreuilly2020}. Using this approximation, we expect the hydrodynamical drift velocity to be approximately proportional to the stopping time of the larger grain, the species $i$. As the stopping time increases with grain size, the hydrodynamical drift should be efficient at forming large grains, more than at removing small grains. Still, as the drift velocity roughly scales as $\propto \frac{1}{\rho^2}$ during the protostellar collapse, the growth by gas-dust drift is probably rapidly suppressed during the collapse unless large grains are assumed in the initial conditions.
\subsubsection{Brownian motion}
\label{sec:brow}
Brownian motions can also generate differential motions between particles. For two grains $i$ and $j$, the differential velocity associated to brownian motions can be written as 
\begin{equation}
\Delta  v_{\mathrm{brownian},i,j} \equiv \sqrt{\frac{8 k_{\mathrm{B}} T}{\pi}} \sqrt{\frac{m_i + m_j}{m_i m_j} }.
\end{equation}
Considering the case where $m_j \ll m_i$, we see that $\Delta  v_{\mathrm{brownian},i,j} \propto \frac{1}{\sqrt{m_j}}$, yielding $K_{i,j}\propto \frac{m_i^{2/3}}{\sqrt{m_j}}$: the coagulation kernel diverges for very small masses (still when $m_j \ll m_i$). We thus expect the growth by brownian motions to be most efficient for small particles colliding with larger particles. 

\subsubsection{Ambipolar diffusion}
The last drift mechanism considered in this study is caused by ambipolar diffusion. It bears resemblance with the hydrodynamical drift as it is due to a decoupling between dust and gas particles. However, contrary to the hydrodynamical drift, the ambipolar drift is due to the coupling of charged dust grains to the magnetic field. It is not simple to estimate the ambipolar diffusion velocity as it should, in principle, be accounted for in the drift velocity via a full treatment, possibly in 3D, of the magnetic field. However we can use the simplified approach proposed by \cite{Guillet2020} to take it into account. The ambipolar diffusion velocity $\vec{V}_{\mathrm{AD}}$ can be written as
\begin{equation}
\vec V_{\mathrm{AD}} \simeq \frac{1}{|\vec{B}|^2}\frac{c^2 \eta_{\rm{AD}} ((\nabla \times\vec{B} )\times \vec{B})}{4 \pi},
\end{equation}
where $c$ is the speed of light, $\eta_{\rm{AD}}$ is the ambipolar resistivity (in unit $s$). As in 
\cite{Guillet2020}, we choose to model the electric current $\nabla \times\vec{B}$ in a simple way by assuming that it is approximately $\frac{|\vec{B}|}{\lambda_J}$ where $\lambda_J$ is the Jeans length. We thus have 
\begin{equation}
V_{\mathrm{AD}} \simeq  \delta \frac{c^2 \eta_{\rm{AD}}}{4 \pi \lambda_J},
\end{equation}
where $\delta$ is close to unity (unless specified we considered $\delta=1$).
Defining the Hall factor $\Gamma_k$ as the ratio of the grain stopping time to its gyration time, we can express the ambipolar drift between the grain $i$ and $j$ as \citep{Guillet2020}
\begin{equation}
\Delta  v_{\mathrm{ambipolar},i,j} \equiv  V_{\mathrm{AD}} \left|\frac{|\Gamma_i| }{\sqrt{1+\Gamma_{i}^2}}-\frac{|\Gamma_j| }{\sqrt{1+\Gamma_j^2}}\right|.
\end{equation}
 It is interesting to point out that this drift, like $\eta_{\rm{AD}}$, increases with a stronger magnetic field. As proposed by \cite{Silsbee2020} and \cite{Guillet2020}, ambipolar diffusion is probably very efficient at removing the smallest grains, which are well coupled to the magnetic field, by sticking them onto large grains (those being more coupled to the gas). However, ambipolar diffusion is unable to generate strong relative velocities between large grains, and is therefore inefficient at making up large grains.

\subsection{Evolution of the dust size distribution}
\label{sec:evolution}
The acceleration and decoupling of grains induced by the various mechanisms described in Sect. \ref{sec:drifts} generate grain-grain collisions that can lead to the coagulation or to the fragmentation of the colliding grains.

For simplicity, we model dust grains as compact spheres. The outcome of any coagulation or fragmentation process following the collision of two grains will lead to the formation of grains that are also spherical and compact.

\subsubsection{Coagulation}
The dust coagulation source term (without fragmentation), $S_{k,\mathrm{growth}}$ (in Eq.~ \ref{eq:hydro}), can be written in its discrete form as \citep{Smolu16}
\begin{equation}
S_{k,\mathrm{growth}}=  \sum_{i+j\rightarrow k} K_{i,j} (m_i+m_j)\, n_{j} n_i -n_k m_k \sum_i^{\mathcal{N}} K_{k,i} n_i,
\end{equation}
where $K$ is the coagulation kernel, the notation $i+j\rightarrow k$ indicates that the summation is over all the binary collisions (counted only once with the convention $j \le i$) that lead to mass added to the bin $k$, $n_i$ is the number density of grains of mass $m_i$ such as $\rho_i \equiv m_i \,n_i$. 

The coagulation kernel is expressed as
\begin{equation}
K_{i,j} = \sqrt{\frac{8}{3\pi}} \pi (s_{i} + s_{j})^2 \Delta  v_{i,j},
\end{equation}
 where $ \Delta  v_{i,j}$ is the differential velocity between grains $i$ and $j$, of size $s_i$ and $s_j$, defined as the quadratic sum of the four sources of differential velocity detailed in Sect. \ref{sec:drifts}. The $\sqrt{\frac{8}{3\pi}}$ pre-factor comes from considering that grains relative velocities along the three x, y and z axis are Gaussian variables of variance $\Delta v_{i,j}^2/3$ \citep{Guillet2020,Marchand2021}.

\subsubsection{Fragmentation}
\label{sec:fragtheo}
Collisions between two grains do not always lead to sticking. 
Other phenomena such as bouncing, fragmentation, compaction, cratering, and erosion also happen \citep[for a summary of the many outcomes of grain-grain collisions, see][]{Guttler2010}.
For simplicity, we limit our modeling to the key mechanisms that will dominate the evolution of the dust size distribution and magnetic resistivities, namely the fragmentation and coagulation. We emphasize that this does not mean that other mechanisms are negligible. Instead they should be investigated in dedicated works.

The microphysics of fragmentation adapted to our study of quiescent environments like protostellar cores is not that of solids breaking up into pieces when colliding at supersonic velocities \citep{Tielens1994,Jones1996,Guillet2011}, but that of porous aggregates composed of monomers and colliding at subsonic velocities \citep{Dominik1997,Ormel2009}. The detailed physics is also quite different: fragmentation occurs above a velocity threshold for the former, and above an energy threshold for the latter.

Let us consider a collision between grain sizes $i$ and $j$. As per \cite{Ormel2009}, we consider that a collision where the projectile and the target are composed of a total $N_{\mathrm{tot}}$ monomer of uniform radius $s_{\rm mono}$ will lead to a complete fragmentation of the colliding mass if 
\begin{align}
E_{\mathrm{kin},i,j} \equiv \frac{1}{2} \frac{m_im_j}{m_i+m_j}  \Delta  v_{i,j}^2 > 5 N_{\mathrm{tot}}  E_{\mathrm{br}}, \\
E_{\mathrm{br}} = A_{\mathrm{br}} \gamma_{\mathrm{grain}}^{5/3}\frac{(s_{\mathrm{mono}}/2)^{4/3}}{\varepsilon^{2/3}},
\end{align}
where $E_{\mathrm{kin},i,j}$ is the kinetic energy of the collision between $i$ and $j$, $E_{\mathrm{br}}$ is the energy required to break a bond between two monomers, and $\varepsilon$ and $\gamma_{\mathrm{grain}}$ are the reduced elastic modulus and surface energy density of the material.
We considered values given by \cite{Ormel2009} both for icy grains and for bare silicates. The latter gives the lowest fragmentation energy threshold, while the former leads to almost insignificant fragmentation \citep[also confirmed by our study][]{Kawasaki2022}. As in   \cite{Ormel2009}, $A_{\mathrm{br}}=2.8 \times 10^{3}$.

The fragmentation energy threshold is easily translated into a velocity threshold $v_{\mathrm{frag},i,j}$
\begin{equation}
v_{\mathrm{frag},i,j}= \ (1 +1/ \mu_{i,j})\sqrt{\frac{ 10 ~\mu_{i,j} E_{\mathrm{br}}}{m_{\mathrm{mono}}}},
\end{equation}
where $m_{\rm mono}$ is the mass of the monomer that compose both aggregates. We stress that this velocity threshold depends on the mass ratio $\mu_{i,j}=m_i/m_j > 1$ of the colliding grains, and is therefore not constant. This velocity threshold is minimized for collisions of equal size grains, and diverge as $\sqrt{m_i/m_j}$ when $m_j \ll m_i$.
For equal size grains, the fragmentation velocity $v_{\mathrm{frag}}$ is then
\begin{equation}
v_{\mathrm{frag}}= \sqrt{\frac{20  E_{\mathrm{br}}}{m_{\mathrm{mono}}}}.
\end{equation}
 Considering that the grains are bare-silicate or ice-coated and assuming $0.1\micro\meter$ monomers \citep{Ormel2009} allows us to determine that $v_{\mathrm{frag}} \sim~ 15  ~\meter \,\second^{-1}$ for silicate grains and  $v_{\mathrm{frag}} \sim 300~ \meter \,\second^{-1} $  for ice-coated grains. These values differ by more than an order of magnitude. In the case of icy grains, the fragmentation is happening at larger grains sizes than for bare-silicates, which was already clear from \cite{Ormel2009}. We emphasize that $v_{\mathrm{frag}}$ is the correct threshold only for a collision of equal size grains.
 
In the context of fragmentation, the expression of $S_{k,\mathrm{growth}}$ is modified and requires to define $f_{\mathrm{frag},i,j}$ the fraction of the colliding mass that goes into fragment and $\alpha_{k,i,j}$ is the distribution of fragments. We now have 
\begin{align}
S_{k,\mathrm{growth}}&= \sum_{i+j\rightarrow k} K_{i,j} (1-f_{\mathrm{frag},i,j})\,(m_i+m_j) n_{j} n_i
\nonumber\\
&+\sum_{i+j\rightarrow k} \alpha_{k,i,j} K_{i,j} f_{\mathrm{frag},i,j}\,(m_i+m_j)  n_{j} n_i
\nonumber\\
&-n_k m_k \sum_i^{\mathcal{N}} K_{k,i} n_i,
\end{align}
 Note that, in order to ensure mass conservation, $\alpha_{k,i,j}$ must verify $\sum_k \alpha_{k,i,j}=1$.
\section{Numerical methods}
\label{sec:method}
\subsection{General presentation of the code}

For the purpose of this work, we developed the {\ttfamily shark} code. It is a 1D finite-volume code that solves the previously introduced equations of hydrodynamics for gas and dust mixture in a spherical geometry. It accounts for the dynamics and growth/fragmentation of a distribution composed of multiple dust species. The code also computes the charging of dust grains, the ion and electron density and the magnetic conductivities and resistivities using the fast ionisation scheme of \cite{Marchand2021}.
{\ttfamily shark} solves the equations of hydrodynamics using the Godunov scheme with a Lax-Friedrich method to estimate the hydrodynamical fluxes. The hydrodynamics equations are spherically averaged and solved as in \cite{Hennebelle2021} (dust drag excluded and without the turbulent terms). We note that a particularity of {\ttfamily shark} is that it solves the equations for $\mathcal{N}+1$ species and that the dust, contrary to the gas, does not feel any pressure force.

{\ttfamily shark} is a multi-purpose tool that can be used to investigate dust evolution in other astrophysical environments than the protostellar collapse e.g., photo-dissociative regions, protoplanetary disks, molecular clouds. 
\subsection{Magnetic field}
Although we do not explicitly solve the induction equation in this work, we still need to estimate the magnetic field strength to compute the resistivities. 
The magnetic field in the regions of low ambipolar diffusion is estimated as in \cite{Marchand2016}, we have
\begin{equation}
B = B_0 \sqrt{ \frac{n_{\mathrm{H}}}{10^{4} \centi\meter^{-3}}}.
\end{equation}
 Unless specified $B_0 = 30~\micro\rm{G}$.  Let us now define the ambipolar diffusion timescale $t_{\mathrm{AD}}$ such as 
\begin{equation}
t_{\mathrm{AD}} =\frac{4\pi}{c^2}\frac{r^2}{\eta_{AD}}.
\end{equation}

We have two possible scaling for the magnetic field. Either the diffusion is very inefficient and $B \propto \sqrt{n}$ or $t_{\mathrm{AD}}<t_{\mathrm{ff}}$ and then $B$ is a constant with respect to the density. We verified for a simple test with a MRN distribution (run $\textsc{PC0}$, see Tab.\ref{tab:models}) that this approach reproduces very well the $0.1~$G plateau typically observed in 3D non-ideal MHD simulations of protostellar collapse \citep{Masson2016}.

\subsection{Dust}
\subsubsection{Dynamics}
In addition from the gas, dust is distributed between $\mathcal{N}$ species. All the dust species are treated as separate fluids interacting with the gas via the drag force. This drag is accounted for using the same implicit scheme as presented by \cite{LLambay2019} and more particularly the updated version proposed by \cite{Krapp2020}. It allows for a fast and unconditionally stable treatment of the gas and dust coupling and does not require any small Stokes approximation \citep{LP2014a,Lebreuilly2019}. The velocity of the dust fluid obtained using this scheme is the one that we used to compute the hydrodynamical drift presented before. The other sources of relative velocity are computed using the expressions that we presented in Sec. \ref{sec:drifts}. 
\subsubsection{Ambipolar drift}
The magnetic field is used to compute the resistivities and the ambipolar drift. To estimate the latter, as described in  Sec. \ref{sec:drifts}, we compute the Jeans length as
\begin{equation}
\lambda_{\mathrm{}} = c_{\mathrm{s}} t_{\mathrm{ff}},
\end{equation}
 Finally, the Hall factor is the one derived from the charge solver as described in \cite{Marchand2021}.
\subsubsection{Distribution}

\begin{table}
       \caption{The dust distributions explored in this work.}      
\label{tab:dust-distri}      
\centering          
\begin{tabular}{c c c c}   
 \hline \hline 

                   MRN & $s_{\mathrm{min,init}}$ [cm] & $s_{\mathrm{max,init}}$ [cm] & $\lambda$  \\
           
                     & $5 \times 10^{-7}$  &  $2.5 \times 10^{-5}$  & -3.5\\ \hline  \hline \\

 \hline  \hline
LOGN&  $s_{\mathrm{mean}}$ [cm] & $\sigma$ \\
& $10^{-4}$ & 1 \\ \hline  \hline
\\ \hline  \hline
 GROWN & ${\nh}_0$ [$\centi\meter^{-3}]$& $t_0$ [Myr]\\
 & $10^5$ & 1  \\

   \hline \hline 
\end{tabular}
\end{table}
The dust density is computed assuming a total dust-to-gas ratio $\epsilon_0=0.01$ and is distributed according to the grain size $s$ on a logarithmic grid that ranges between $s_{\mathrm{min}}$ and $s_{\mathrm{max}}$. Let us define the logarithm increment $\zeta \equiv (s_{\mathrm{max}}/s_{\mathrm{min}})^{1/\mathcal{N}}$. For each bin $i$,  grain size is comprised between $s_{-,i}\equiv s_{\mathrm{min}}\, \zeta^{i-1}$ and $s_{+,i}\equiv s_{\mathrm{min}}\, \zeta^{i}$ and we choose $s_{i}=\sqrt{s_{-,i} s_{+,i}} + s_{\mathrm{ices}}$ as the typical bin size. $s_{\mathrm{ice}}= 8.7~\nano\meter$ is the size of the ice mantle on the grains of which we neglect the mass. Note that $s_{i}$ is the size that is used by the solver to compute the stopping time and therefore the dynamics. We also define the grain masses as $m_{-,i} =\frac{4}{3} \pi \rho_{\mathrm{grain}} s_{-,i}^3$, $m_{+,i} =\frac{4}{3} \pi \rho_{\mathrm{grain}} s_{+,i}^3$ and $m_{i} =\sqrt{m_{-,i} m_{+,i}} $, where $\rho_{\mathrm{grain}}$ is the grain intrinsic density ($\rho_{\mathrm{grain}}=2.3 \gcc$) . In Tab.~\ref{tab:dust-distri}, we describe the initial dust size distributions explored in this work. A complete description of their modeling is detailed in Appendix A.

\subsubsection{Growth}

\begin{figure}
\centering
     \includegraphics[width=
          0.4\textwidth]{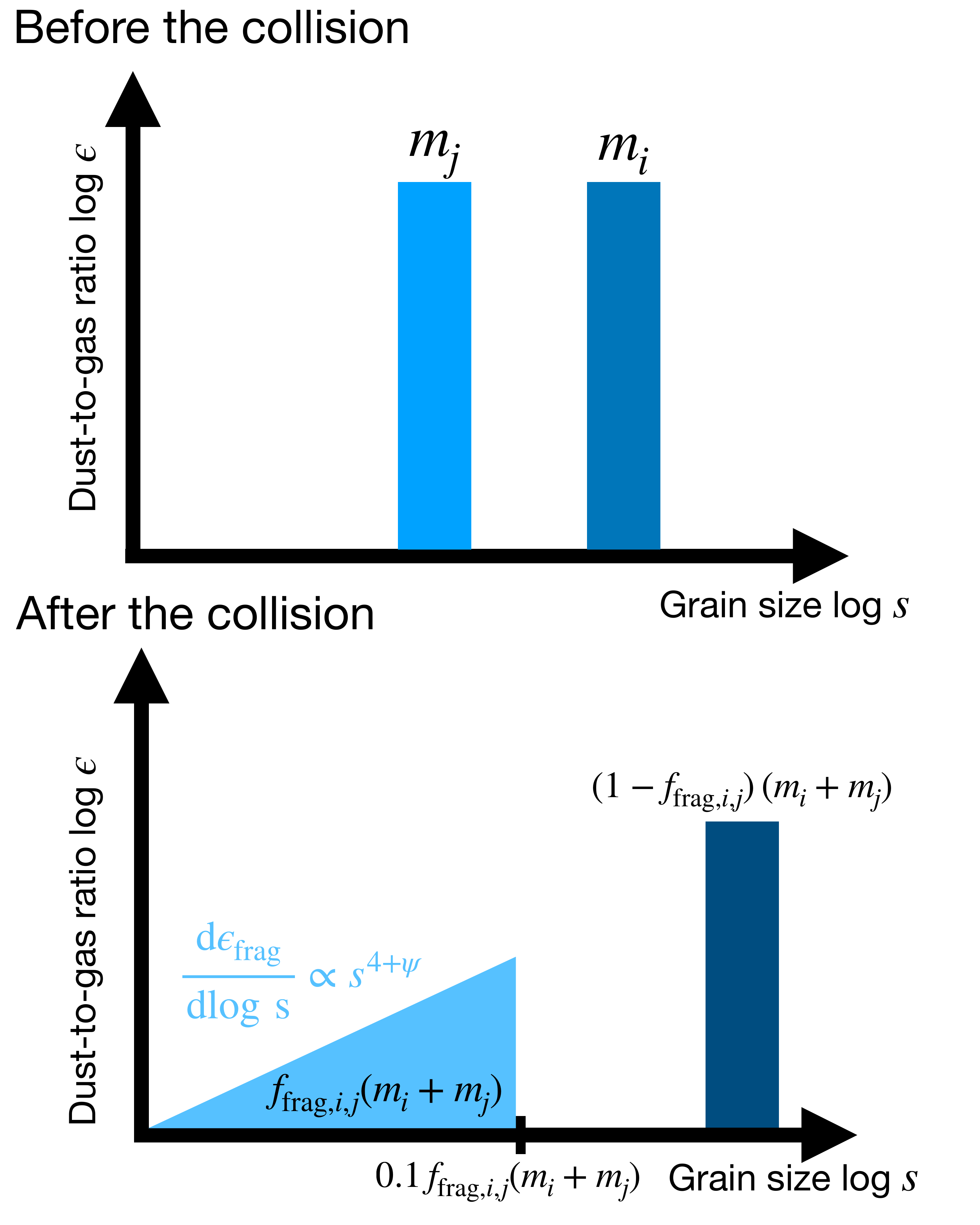}
     \caption{Cartoon illustration explaining the outcome of a collision in our model. (Top): Before the collision  between $i$ and $j$. (Bottom): After the collision, a main fragment of mass $(1-f_{\mathrm{frag},i,j})\, ( m_i +m_j)$ is formed ($0<f_{\mathrm{frag},i,j}<1$) and the remaining of the mass is distributed as a power law up to the bin of mass $0.1 f_{\mathrm{frag},i,j}\, ( m_i +m_j)$.  }
            \label{fig:coag_scheme} 
\end{figure}

The dust growth is solved according to a scheme similar to those presented by \cite{Guillet2007}, \cite{Guillet2020} and \cite{Marchand2021}. For two bins $i$ and $j\le i$, we compute the coagulation rate as  
\begin{equation}
  \left(\frac{\mathrm{d} n}{\mathrm{d} t}\right)_{i,j} =  n_i n_j K_{i j},
\end{equation}
   this rate is divided by two when $i=j$ as to count the collisions of equal size grains only once.  
 
In our model, a collision  can simultaneously lead to  the fragmentation and coagulation of colliding grains. A mass $f_{\mathrm{frag},i,j}\, (m_i +m_j)$ of the colliding grains will go into a size distribution of fragments, and the rest of the mass $(1-f_{\mathrm{frag},i,j})\, ( m_i +m_j)$ will form a new grain that will feed into the collector bin $k_{i,j}$ that verifies $m_{-,k_{i,j}} \le (1-f_{\mathrm{frag},i,j})\, ( m_i +m_j) < m_{+,k_{i,j}}$. We point out that, contrary to \cite{KovetzOlund1969,Brauer2008}, we consider a single collector bin, thus our method relies on using a large number of dust bins to converge. Allowing the coagulated mass to be shared with two coagulated bins, as is proposed in the appendix A1 of \cite{Brauer2008}, would be a solution to converge better with fewer dust bins.  Naturally, $f_{\mathrm{frag},i,j}=0$ when fragmentation is not included in the model.

Once we determined $\left(\frac{\mathrm{d} n}{\mathrm{d} t}\right)_{i,j}$ for each pair of $i$, $j\le i$ we can compute the coagulation rates
\begin{align}
\left(\frac{\mathrm{d} \rho_i}{\mathrm{d} t}\right)_{i,j}  &= -  \, m_{i} \left(\frac{\mathrm{d} n}{\mathrm{d} t}\right)_{i,j} \nonumber \\
\left(\frac{\mathrm{d} \rho_j}{\mathrm{d} t}\right)_{i,j} &= -   \, m_{j} \left(\frac{\mathrm{d} n}{\mathrm{d} t}\right)_{i,j} \nonumber \\
\left(\frac{\mathrm{d} \rho_{k_{i,j}}}{\mathrm{d} t}\right)_{i,j} &=  (1-f_{\mathrm{frag},i,j}) \, (m_{i}+m_{j}) \left(\frac{\mathrm{d} n}{\mathrm{d} t}\right)_{i,j}.
\end{align}
We redistribute the mass of fragments in a spectrum of grain sizes. In our model, we assume as per \cite{Kawasaki2022} that fragments are redistributed in a power law size distribution of constant power law index $\psi$, with grain sizes ranging from $s_{\rm min}$ up to the size grains of mass $0.1f_{\mathrm{frag},i,j} \left( m_{i}+  m_{j}\right)$ contained in bin $l_{i,j}$. The mass transfer rate in bin $l$ is 

\begin{equation}
\left(\frac{\mathrm{d} \rho_{l}}{\mathrm{d} t}\right)_{i,j} =  f_{\mathrm{frag},i,j} \alpha_{l,i,j} \left( m_{i}+  m_{j}\right)  \left(\frac{\mathrm{d} n}{\mathrm{d} t}\right)_{i,j},
\end{equation}
where
\begin{equation}
\alpha_{l,i,j} = \left\{
    \begin{array}{ll}
       \left(s_{l,+}^{4+\psi}-s_{l,-}^{4+\psi}\right)/\left(s_{l_{i,j},+}^{4+\psi}-s_{\mathrm{min}}^{4+\psi}\right) & \mbox{if }l\le l_{i,j} , \\
        0 & \mbox{otherwise,}
    \end{array}
\right.
\label{eq:ak}
\end{equation}
Mass conservation is guaranteed since $\sum_{l=1}^{\mathcal{N}} \alpha_{l,i,j}=1$.  We show, in Fig.~\ref{fig:coag_scheme}, a cartoon illustrating the result of a binary collision.

We sum the mass transfer rates for all the possible collisions to get a total rate $\left(\frac{\mathrm{d} \rho_i}{\mathrm{d} t}\right)$ for each bin $i$ that we use to update the density $\rho_i$ according to 
\begin{equation}
    \rho_i^{\mathrm{new}} = \rho_i^{\mathrm{old}} + \Delta t_{\mathrm{growth}} \left(\frac{\mathrm{d} \rho_i}{\mathrm{d} t}\right),
\end{equation}
where $\Delta t_{\mathrm{growth}} =  \mathrm{min}_{ i \in \mathcal{N}}(C \rho_i^{old}/\left(\frac{\mathrm{d} \rho_i}{\mathrm{d} t}\right))$ with $C=0.5<1$ is computed to make sure that no negative densities are obtained. Let us define $\Delta t$ the simulation timestep (computed according to the Courant-Friedrich-Lewy (CFL) condition \citep{Courant1928}, there are two possibilities, either  $\Delta t< \Delta t_{\mathrm{growth}}$ and thus we consider $\Delta t_{\mathrm{growth}} =\Delta t $, or we sub-cycle the dust growth by doing several individual timesteps (using the previous constraint) until the sum of these individual timesteps is $\Delta t$. As not all the physical cells have a very constraining stability conditions we choose to sub-cycle the growth cell-by-cell to minimise as much as possible the time spent in the growth solver. Note that even then, the computational time is mostly spent in the growth algorithm. 

The growth algorithm is tested against analytical solutions in Appendix B. 

\subsubsection{Fragmentation recipe}

For the numerical implementation of fragmentation, we base our expression for $f_{\mathrm{frag},i,j}$ on Fig.~5 from \cite{Ormel2009}. The mass  is approached by the following function
\begin{equation}
f_{\mathrm{frag},i,j}=  \left\{
    \begin{array}{ll}
       1  & \mbox{if } E_{\mathrm{kin},i,j} > 5 N_{\mathrm{tot}}  E_{\mathrm{br}} , \\
       0 & \mbox{ if } E_{\mathrm{kin},i,j} <  0.1\, N_{\mathrm{tot}} E_{\mathrm{br}} \\
          \propto E_{\mathrm{kin},i,j} & \mbox{otherwise.}  
    \end{array}
\right.
\end{equation}

If $E_{\mathrm{kin},i,j} > 5 N_{\mathrm{tot}}  E_{\mathrm{br}}$, the grains will break entirely into fragments. If $E_{\mathrm{kin},i,j} < 0.1 N_{\mathrm{tot}}  E_{\mathrm{br}}$, the collision will lead to the total  coagulation of the projectile and target grains. In between, partial coagulation and partial fragmentation happens.

\subsubsection{Charging and resistivities}
We use the method from \cite{Marchand2021} to compute the ionisation state of the cloud. A test of the scheme and a small improvement of the method can be found in the Appendix B. For more information, the method was extensively presented in \cite{Marchand2021}. The ionisation and the values of the resistivities depend on several parameters like the cosmic-rays ionisation rate $\zeta_{\mathrm{CR}}$, the average ion mass $\mu_{\mathrm{ions}}$, the size of the ice mantles $s_{\mathrm{ices}}$ on the grains and the sticking efficiency of electrons onto grains $s_{\mathrm{e}}$. In this paper, we have chosen the same values as those employed by \cite{Marchand2021}, i.e. $\zeta_{\mathrm{CR}} = 5 \times 10^{-17}~\second^{-1}$, $\mu_{\mathrm{ions}}=25$ and  $s_{\mathrm{e}}=0.5$, but we emphasize that they are free parameters in the code.

The conductivities of each species $\sigma_j$ (ions and electrons included) are computed the same way as in  \cite{Marchand2021}. We recall here their definition. We define the parallel, perpendicular and Hall conductivities: $\sigma_{\mathrm{par}}$, $\sigma_{\mathrm{perp}}$ and $\sigma_{\mathrm{H}}$, that can be computed from the conductivities of all the individual charged species $\sigma_j$ such as
\begin{align}
\sigma_{\mathrm{par}}&= \sum_j \sigma_j, \nonumber \\
\sigma_{\mathrm{perp}}&=  \sum_j  \sigma_j \frac{1}{1+\Gamma_j^2} , \nonumber \\
\sigma_{\mathrm{H}}&=- \sum_j  \sigma_j \frac{\Gamma_j}{1+\Gamma_j^2}.
\end{align}
Since all the individual conductivities are positive and since $\Gamma_j$ can be negative, it is quite clear that the Hall conductivity can change sign depending on the dominant charge carrier.  

Once the conductivities are known, the Ohm, ambipolar and Hall resistivities, $\eta_{\mathrm{O}}$, $\eta_{\mathrm{AD}}$ and $\eta_{\mathrm{H}}$ are simply computed as 
\begin{align}
\eta_{\mathrm{O}}&= \frac{1}{\sigma_{\mathrm{par}}}, \nonumber \\
\eta_{\mathrm{AD}}&=\frac{\sigma_{\mathrm{perp}}}{\sigma_{\mathrm{perp}}^2+\sigma_{\mathrm{H}}^2} -\frac{1}{\sigma_{\mathrm{par}}} , \nonumber \\
\eta_{\mathrm{H}}&= \frac{\sigma_{\mathrm{H}}}{\sigma_{\mathrm{perp}}^2+\sigma_{\mathrm{H}}^2}.
\end{align}
It is again clear that the Hall resistivity can change sign, in fact it holds the same sign as $\sigma_{\mathrm{H}}$.

\subsection{Setup}

\label{sec:setup}
We initialize our clouds as spheres of uniform density according to the \cite{Boss1979} setup. Given an initial gas mass $ M_{0} = 1 M_{\odot}$, its initial radius $R_0$ is set according to the thermal-to-gravitational energy ratio
\begin{equation}
\alpha \equiv \frac{5}{2} \frac{R_0 k_{\rm{B}} T_{0}}{\mathcal{G} M_0\mu_{\rm{g}}m_{\rm{H}}},
\end{equation}
where $ T_{0}=10~$K is the initial cloud temperature. In this work, we investigate clouds with $\alpha = 0.25$, which leads to a radius $R_0\sim 2500~$au  and initial density $\rho_0=9.2 \times 10^{-18}~\gcc$.

To model the thermal evolution of the cloud in an appropriate way, in the context of collapse simulations, we compute the pressure according to a barotropic equation of state
\begin{equation}
    P = C_{\mathrm{s},0}^2 \left(1+\left[\frac{\nh}{n_1}\right]^{0.8}\right)^{\frac{1}{2}}\left(1+\left[\frac{\nh}{n_2}\right]\right)^{-0.3}\left(1+\left[\frac{\nh}{n_3}\right]\right)^{-\frac{1.7}{3}},
\end{equation}
where $n_1=10^{11}~ \centi\meter^{-3}$, $n_2=10^{16}~ \centi\meter^{-3}$, $n_3=10^{21}~ \centi\meter^{-3}$, and $C_{\mathrm{s},0} \equiv \sqrt{\frac{\gamma k_{\mathrm{B}}T_0}{\mu_{\rm{g} m_{\rm{H}}}}}$ is the gas sound speed at low densities. This is the same equation of state as used in the previous studies of \cite{Machida2006,Marchand2016,Marchand2021}.

{\ttfamily shark} can handle non-regularly spaced grid which is convenient for collapse calculations. In our models we use a logarithmic grid composed of $\mathcal{N}_{\mathrm{cells}}$ cells with a radius ranging between $r_{\mathrm{in}} =1 ~\mathrm{au}$ and $r_{\mathrm{max}}= R_0$. The inner and outer boundary conditions are zero density gradient. In addition, the inner boundary has a zero velocity for both the gas and the dust. We have chosen to use $\mathcal{N}_{\mathrm{cells}}=128$, which leads to a spatial resolution of  approximately $\sim 30$ points per decade.

For all the runs, we consider $\mathcal{N}=100$ dust species of sizes between $s_{\mathrm{min}}=7~\nano\meter$ and $s_{\mathrm{max}}=1~\centi\meter$ initialised with the various distributions presented in Appendix~\ref{sec:appendix_distri}. When fragmentation is included, we need to make a choice for the fragment distribution and the monomer properties. We considered that a good first guess was a standard MRN. Hence, we imposed $\psi=-3.5$. We also choose a average monomer size $s_{\mathrm{mono}}\sim 0.1~\micro\meter$ as in 
\cite{Ormel2009}.

\section{Results}
\label{sec:results}
The models computed for this work are summarised in Tab.~\ref{tab:models}. We evolved all the protostellar collapses calculations up to the point when the maximum density reaches $10^{-10}\gcc$, at about $t \sim 20 $~yr after the formation of the first Larson core. We note that we consider the first Larson core to be formed when the density reaches $10^{-11}\gcc$, i.e. at $\sim 21.865~$kyr in our runs. Unless specified (in the case of run $\textsc{PC1-FRAG}$), the models do not include grain fragmentation. To help the comparison between models, we computed $\textsc{PC0}$, which is the same run as $\textsc{PC1}$ but without grain growth. In this model the MRN distribution is extremely well preserved as small grains are very well coupled to the gas.

\begin{table*}
       \caption{Syllabus of the different simulations. From left to right: model name, magnetic field strength at a density of $10^4 \centi\meter^{-3}$, pre-factor of the ambipolar velocity $\delta$,  inclusion of growth, fragmentation, hydrodynamical drift, turbulence, brownian motion and ambipolar diffusion and initial choice of dust distribution.}      
\label{tab:models}      
\centering          
\begin{tabular}{c c c c c c c c c c}     
\hline\hline       
                   Model &  $B_0$ [$\micro$G]& $\delta $&Growth & Fragmentation & Drift & Turbulence & Brownian & Ambipolar &   Distribution  \\
\hline      
$\textsc{PC1-FRAG}$    & 30  & 1 &  \checkmark &   \checkmark  & \checkmark & \checkmark & \checkmark & \checkmark & MRN \\
  $\textsc{PC1}$& 30   & 1 & \checkmark & - & \checkmark & \checkmark & \checkmark & \checkmark & MRN  \\
    $\textsc{PC1-weakB}$& 15   & 1  &\checkmark & - & \checkmark & \checkmark & \checkmark & \checkmark & MRN  \\
     $\textsc{PC1-strongB}$& 50   & 1 & \checkmark & - & \checkmark & \checkmark & \checkmark & \checkmark & MRN  \\
       $\textsc{PC1-weakAD}$& 30   & 0.1 & \checkmark & - & \checkmark & \checkmark & \checkmark & \checkmark & MRN  \\
 \hline 
$\textsc{PC1-GROWN}$ &30   & 1  & \checkmark & - & \checkmark & \checkmark & \checkmark & \checkmark & GROWN \\
$\textsc{PC1-LOGN}$   &30    & 1  &\checkmark & - & \checkmark & \checkmark & \checkmark &  \checkmark &LOGN \\
 \hline 
  $\textsc{PC1-DRIFT}$   & -  & 1 & \checkmark & - & \checkmark & - & - &  - &MRN \\
$\textsc{PC1-TURB}$    & - & 1 &\checkmark & - & - & \checkmark & - &  - & MRN  \\
$\textsc{PC1-BROW}$ & - & 1 & \checkmark & - & - & - & \checkmark & - &  MRN \\
$\textsc{PC1-AMBI}$  & 30  & 1 & \checkmark & - & - & - & - &\checkmark &  MRN \\
  $\textsc{PC1-TURBBROW}$& -   & 1 & \checkmark & - & - & \checkmark & \checkmark &- & MRN  \\
$\textsc{PC1-TURBAD}$& 30   & 1 & \checkmark & - & - & \checkmark &- & \checkmark & MRN  \\
 \hline 
$\textsc{PC0}$    & 30 & 1 & -&   - & - & - & - & - & MRN \\

 \hline \hline
\end{tabular}
\end{table*}

\subsection{Fiducial run}

\begin{figure*}
\centering
     \includegraphics[width=
          \textwidth]{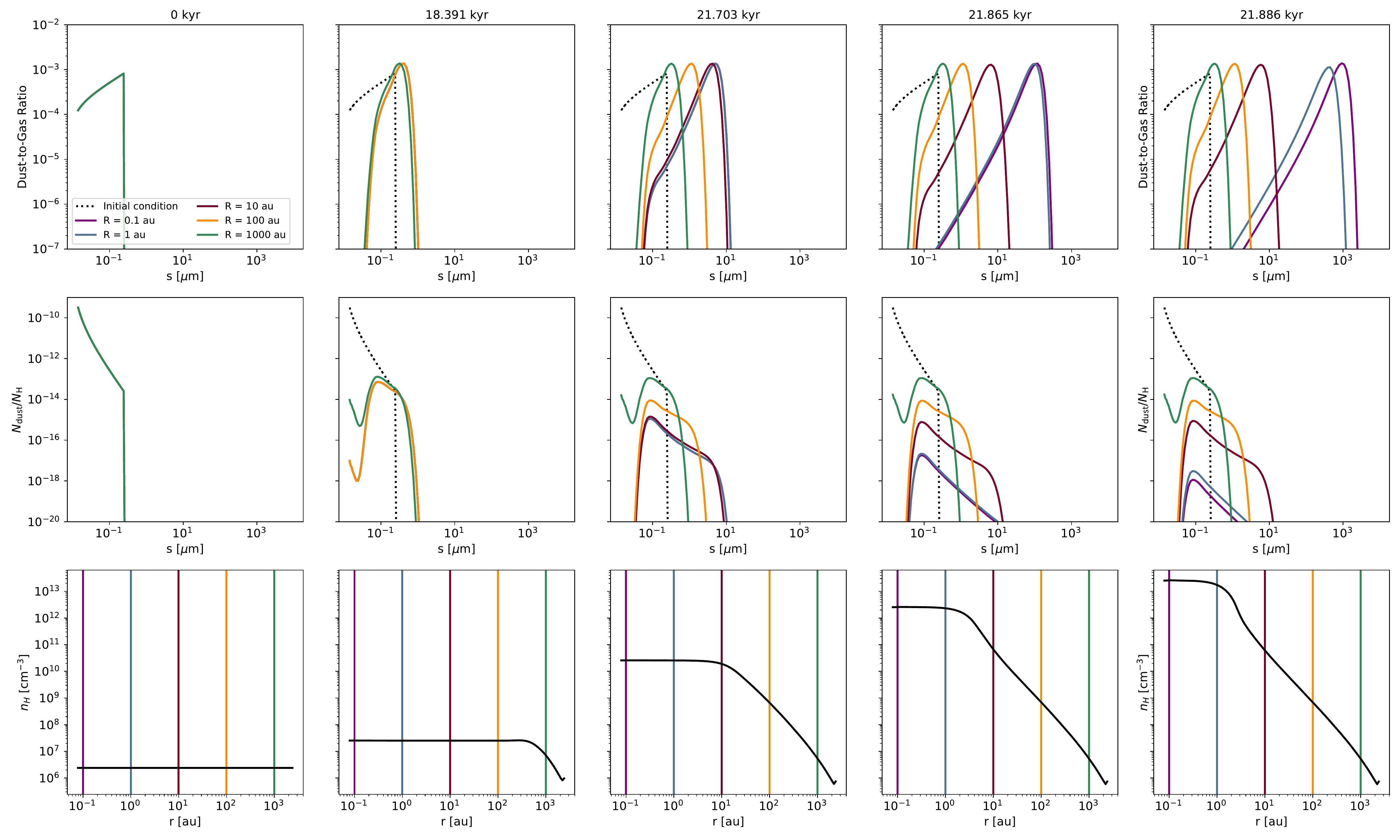}
     \caption{Time evolution (from left to right) of the dust-to-gas mass ratio (top) and number ratio (middle) distribution and the gas density (bottom) for run $\textsc{PC1}$. The four times displayed correspond to the initial condition (left), and then from left to right when the maximal density reaches $10^{-16}\gcc $, $10^{-13}\gcc $, $10^{-11}\gcc$ and $10^{-10}\gcc$, respectively. The vertical coloured lines on the density plots represent the position at which the distribution of corresponding colour are displayed on the top panels. The initial distribution is displayed in dotted lines.}
            \label{fig:time_evol_fid} 

\end{figure*}

Let us introduce our fiducial run $\textsc{PC1}$ that has a standard MRN as initial dust size distribution and includes all the aforementioned source of relative velocities between dust grains. 

Figure \ref{fig:time_evol_fid} shows the time evolution (from the left to the right) of the dust distribution in dust-to-gas mass ratio (top) and dust-to-gas number ratio (middle) and the gas density (bottom) for $\textsc{PC1}$. It is quite clear that for most of the free-fall, the dust does not grow significantly. We can indeed see on the panels of the second column that at $t=18.4~$kyr (when the peak density is around $10^{-16}\gcc $), the peak of the dust distribution is almost at the same position as initially, only shifted by a factor of $< 2$. However small grains have already started to be removed by this time. Contrary to earlier times, there is a significant evolution of the distribution between $t=18.4~$kyr and $t=21.7~$kyr and even more after that. First of all, at $t=21.7~$kyr, the peak of the distribution has shifted to around $1~\micro\meter$ at $r=1$~au and $r=10$~au. At these locations, the distribution is more evolved because the coagulation timescale is shorter at high density. Interestingly, the distribution at high radii has continued to evolve at this stage. This specific behavior cannot be captured by one zone models that follow a collapsing fluid particle through both time and space. Still at $t=21.7~$kyr, the population of small grains has continued to decline everywhere. As shown in previous studies \citep{Silsbee2020,Guillet2020}, and confirmed later in ours, this decline is mostly due to ambipolar diffusion. We complement that brownian motions are also playing a small (but almost negligible compared to ambipolar diffusion) role in that removal of small grains. Finally, an interesting effect is happening around the time of the first core formation, i.e. $t> 21.86~$kyr. We see that in the first core, at $r<5$~au, growth is very efficient. In a matter of only a few years (between  $t= 21.7~$kyr and  $t=21.86~$kyr), the peak has shifted from $\approx 1\micro\meter$ to $\approx 10~\micro\meter$. Later, between $t\simeq 21.86~$kyr and the end of the calculation at $t\simeq 21.9~$kyr, the peak has shifted even more and grains of $\sim 100 ~\micro\meter$ are formed.  Because the density and temperature of the first core are higher, the collision rate between dust grains is boosted and dust growth becomes very efficient. This is consistent with the recent findings of \cite{Bate2022} and \cite{Kawasaki2022}. Additionally, the decline of the number of grains (relative to gas particles) near the end of the collapse is very clear. This is simply due to the extremely large ratio of mass between large and small grains. If the former dominate the mass distribution, they are still vastly outnumbered by small grains. The efficiency at which small grains are removed (ambipolar diffusion) therefore essentially controls the number of grains, while the efficiency at which grains grow to bigger sizes (by turbulence) controls the peak of the mass distribution.
Interestingly, the distribution at $r=1~$au of the last snapshot has a slightly different shape than the distribution at other radii. This is because the differential dynamics between the gas and the dust (the dust drift) is the strongest at the border of the first core. We indeed observe a variation of $\pm 10\%$ of dust-to-gas ratio below and above the accretion shock. Apart from the accretion shock, we note only negligible dust-to-gas ratio variations due to the differential dynamics. This was expected, as shown in dynamical studies \citep{Bate2017,Lebreuilly2019,Lebreuilly2020} strong dust-to-gas ratio variations would only be expected for grains of size larger than a few hundred of microns should they already be present in the initial stages of the collapse, which is not the case here.

\subsection{Impact of the sources of relative velocity}\label{sec:impact_source_velocity}
\begin{figure*}
\centering
     \includegraphics[width=
          \textwidth]{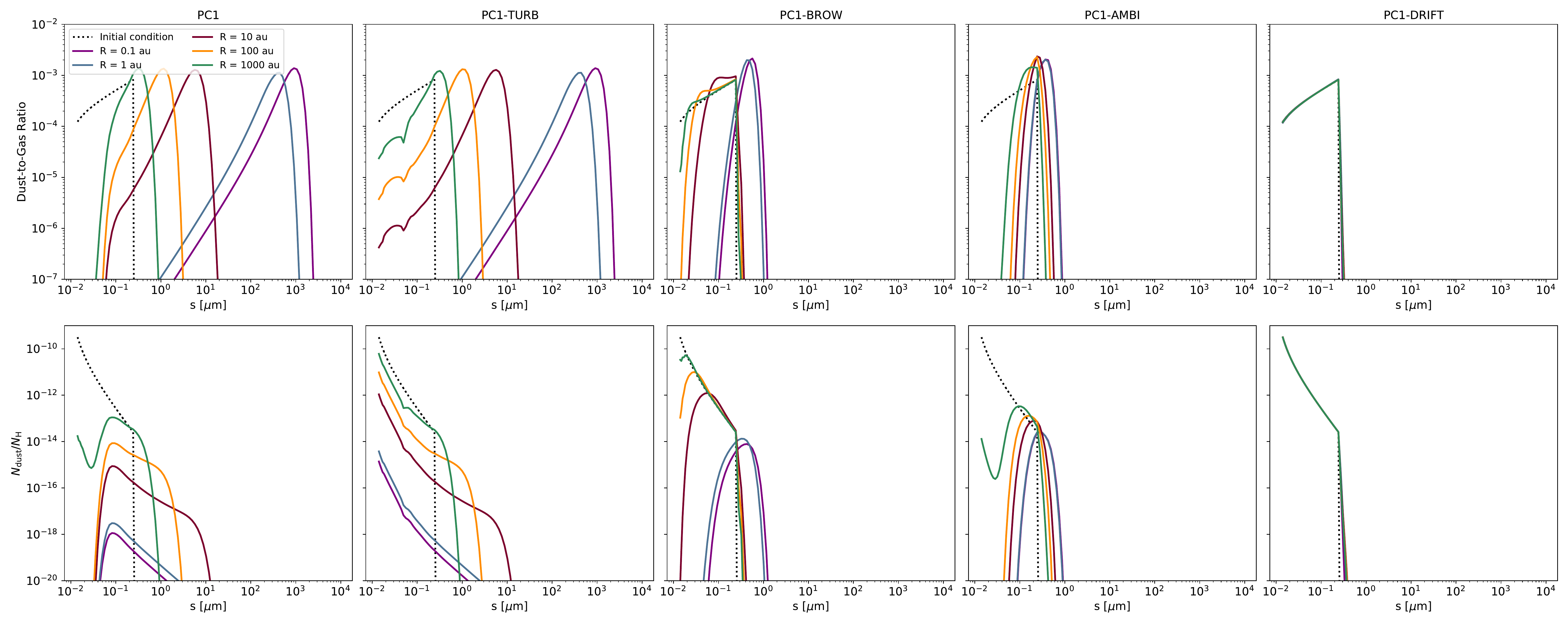}
     \caption{ Final dust distribution of the dust-to-gas ratio mass ratio (top) and the  dust-to-gas number ratio (bottom) for 5 models with different sources of relative velocities. The initial distribution is displayed in dotted lines. From left to right : all sources, turbulence ($\textsc{PC1-TURB}$), brownian motions ($\textsc{PC1-BROW}$), ambipolar diffusion ($\textsc{PC1-AMBI}$) and hydrodynamical drift ( $\textsc{PC1-DRIFT}$). }
                \label{fig:comp_vel} 
    
\end{figure*}
 \begin{figure}
\centering
     \includegraphics[width=
          0.35\textwidth]{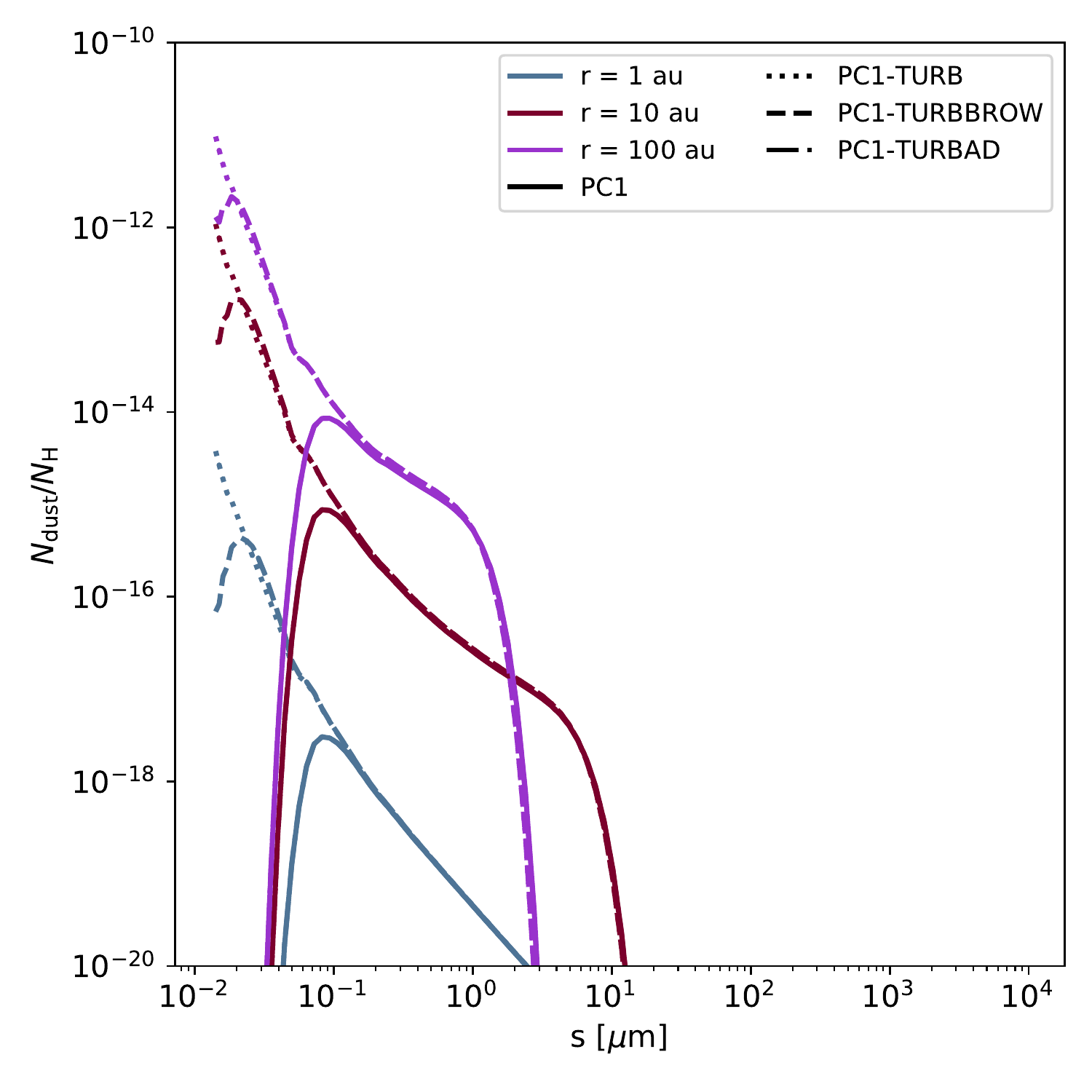}
     \caption{ Final dust number distribution for $\textsc{PC1}$ (plain line), $\textsc{PC1-TURB}$ (dotted line), $\textsc{PC1-TURBBROW}$ (dashed line) and $\textsc{PC1-TURBAD}$ (dot-dashed line) at r = 1 (blue), 10 (red) and 100 (purple) au. }
                \label{fig:comp_smallgrain} 
    
\end{figure}
Let us now investigate the impact of the different sources of relative velocities with four additional models that include only one source each, $\textsc{PC1-DRIFT}$ (hydrodynamical drift only),  $\textsc{PC1-TURB}$ (turbulence only), $\textsc{PC1-BROW}$ (brownian motions only) and $\textsc{PC1-AMBI}$ (ambipolar drift only). In Fig.~\ref{fig:comp_vel} we display the dust-to-gas mass (top) and number (bottom) of all models at the last snapshot of the simulation. As a reference,  we display same information for the $\textsc{PC1}$ run on the left panels. 
 
 It is extremely clear, by looking at the $\textsc{PC1-TURB}$ and $\textsc{PC1}$ panels that turbulence is the main source of large grain formation, and by far. The large grain tails of the distributions of  $\textsc{PC1-TURB}$ and $\textsc{PC1}$ (in mass or in number) are indistinguishable. This same observation was made by the previous works of \cite{Silsbee2020,Guillet2020}. Our distributions are also very similar to those of \cite{Kawasaki2022}. The dominance of turbulence over the other processes is essentially due to the increase of the turbulent velocity with the dust size (in the intermediate coupling regime, i.e. class II vortices) that makes it a very efficient mechanism for collisions of equal size grains. We note that \cite{Bate2022} has found that brownian motions were the dominant mechanism to grow large grains. We think that the difference may arise from the difference in the choice of the Reynolds number. It is indeed set to the fixed value of $10^{8}$ in their study, while in our Eq.~\eqref{eq:Re} \citep{Guillet2020,Marchand2021,Kawasaki2022}, it increases as the square root of the density and temperature and thus reaches larger values as collapse proceeds. The consequence is that, while \cite{Bate2022} mostly considers the tightly coupled regime of turbulence for which the differential velocity of grains of equal sizes is very small, our grains are mostly in the intermediate regime where relative velocities are high. We quite clearly see the change of regime between the class I and II in the panel of  $\textsc{PC1-TURB}$ (at around $0.1\micro\meter$) that can also be seen in \cite{Kawasaki2022}. Similarly to \cite{Bate2022}, we also find that brownian motion can slightly shift the peak of the distribution to about $\sim 1~\micro\meter$ grains and that it leads to an almost monodisperse distribution when acting as a sole process. We did not evolve the model long after the first Larson core formation to focus on the protostellar collapse, but as was pointed out by  \cite{Bate2022}, brownian motion could indeed produce $\sim 100~\micro\meter$ grains in later stages.  
 
 Interestingly, the hydrodynamical drift does not efficiently grow grains. For a standard MRN distribution it is almost completely negligible. As was observed in  \cite{Lebreuilly2020}, the MRN distribution almost does not evolve dynamically during the protostellar collapse, large grains are required to have an important gas-dust drift. Similar findings have been reported in 2D by \cite{Tu2022}. They indeed have shown that the hydrodynamical drift was insufficient to form large mm/cm grains during the Class 0 phase. This does not mean that it is always negligible as this drift also gets stronger with an increasing grain size and most likely plays a role in protoplanetary disks \citep{Birnstiel2009}.

 Let us now focus on the depletion of small grains during the collapse. Comparing $\textsc{PC1-TURB}$ and $\textsc{PC1}$ shows quite obviously that the turbulence is not efficient at depleting small grains. As the grains grow, the distribution is shifted toward larger and larger sizes which causes a decline in the total number of grains (larger grains are less abundant), but this general shift does not provoke a complete depletion of the small grains.  However, not only the ambipolar diffusion, but also the brownian motions \citep[see][for a very similar findings]{HirashitaOkumai2009,Bate2022} are both very effective in that matter. We now focus on $\textsc{PC1-AD}$ and $\textsc{PC1-BROW}$ and the number distribution plots, for which the small grain removal is particularly clear. In the two models, the abundance of small grains relative to the gas severely drops at the inner radii. The brownian motion provokes a significant depletion in the small grains population up to $\simeq 0.1~\micro\meter$ sizes. However, its effect is almost negligible at low density. On the contrary, the small grain removal ($< 0.1~\micro\meter$) is already very effective at a 1000 au scale in the case of $\textsc{PC1-AD}$. This indicates that ambipolar diffusion is faster and more efficient at low densities than brownian motion. In both cases, this small grain removal has important consequences on the total number of dust grains. 
 
 Still, we cannot extrapolate the effect of brownian motion and ambipolar diffusion from their impact as standalone processes. To understand better their respective impact on the population of small grains, we computed  $\textsc{PC1-TURBBROW}$ and  $\textsc{PC1-TURBAD}$, which are the same models as  $\textsc{PC1}$ but with only brownian motions and ambipolar diffusion in addition to turbulence, respectively. We show in Fig.~\ref{fig:comp_smallgrain}  the final dust number distribution for $\textsc{PC1}$ (plain line), $\textsc{PC1-TURB}$ (dotted line), $\textsc{PC1-TURBBROW}$ (dashed line) and $\textsc{PC1-TURBAD}$ (dot-dashed line) at r = 1 (blue), 10 (red) and 100 (purple) au. It is pretty clear that ambipolar diffusion is the most efficient process to reduce the abundance of small grains. The distribution of  $\textsc{PC1-TURBAD}$ and  $\textsc{PC1}$  are indeed indistinguishable. The brownian motion does also reduce the abundance of very small grains but in a much less efficient way. This is not very intuitive since brownian motion is efficient as a standalone process. In the presence of grain growth by turbulence, the total surface area carried by large grains rapidly decreases. Under such conditions, brownian motions become inefficient at removing small grains by sticking them onto large grains, while ambipolar diffusion remains efficient owing to the much larger relative velocities that it generates. We note that ambipolar diffusion may not necessarily be as efficient as we assume. The ambipolar velocity is indeed not very well constrained, which is why we introduced the $\delta$ parameter. We have run an additional calculation with $\delta=0.1$, i.e. with an ambipolar drift ten times smaller, and this strongly reduces the depletion of small grains. However, even in that case, ambipolar diffusion is more efficient than brownian motion at removing small grains.

\subsection{Impact of the initial distribution}

 \begin{figure*}
\centering
     \includegraphics[width=
          0.7\textwidth]{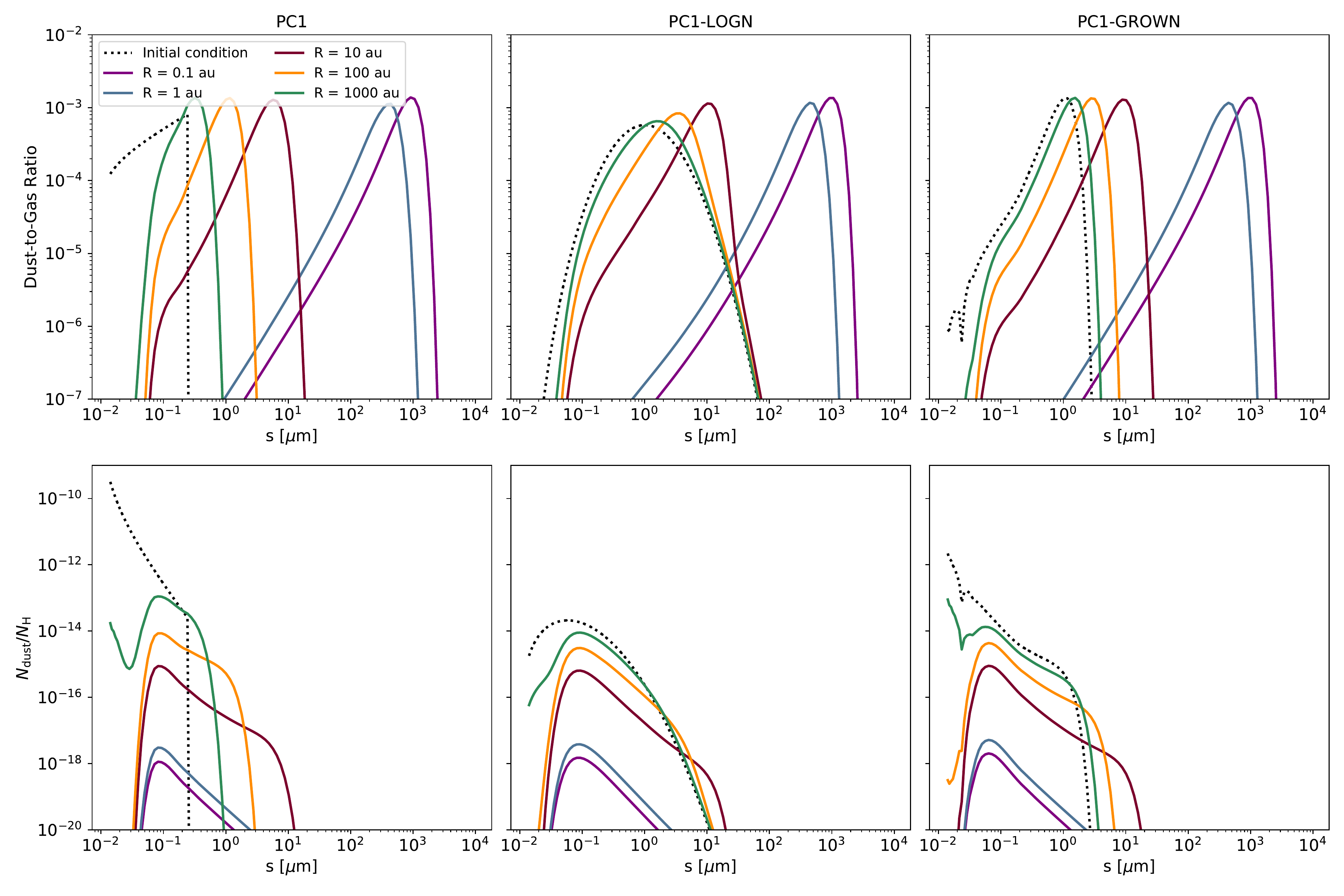}
     \caption{ Final dust distribution of the dust-to-gas ratio mass ratio (top) and the  dust-to-gas number ratio (bottom) for  the 3 models with a different initial dust distribution. The initial distribution is displayed in dotted lines. From left to right : MRN distribution, log-normal distribution and pre-grown distribution. }
                \label{fig:comp_init} 
\end{figure*}
 The validity of the MRN distribution is highly questionable in the context of molecular clouds and during the protostellar collapse \citep{Kohler2015,Jones2013,Jones2017}. To explore the effect of the initial distribution on grain growth, we therefore present two additional runs. Both models have the same conditions as $\textsc{PC1}$ except for the initial choice of dust distribution. In $\textsc{PC1-LOGN}$, we considered a log-normal distribution, which is sometimes claimed to be more realistic than the MRN for the dense interstellar medium dust grains \citep{Jones2013}. In the second run, $\textsc{PC1-GROWN}$, we proposed an alternative approach that assumes grain growth prior to the protostellar collapse (for 1 Myr at $10^{5}~\centi\meter^{-3}$), the conditions at which this distribution has been 'pre-grown' are described in Appendix~\ref{sec:appendix_preg}. In Fig.~\ref{fig:comp_init}, we show the final dust distribution in dust-to-gas mass (top) and number (bottom) ratio for these three models (from left to right). A clear observation can be made: the three models  have an extremely similar dust mass distribution at high density. This is most likely a consequence of the self-similar behavior of the Smoluchowski equation for a wide variety of kernels \citep{Lai1972,Menon2004,Niethammer2014}. However, even if the mass distribution of the three models are similar at high density, this is not the case for a wide range of intermediate densities (and radii).
At low densities, i.e. in the collapsing envelope, the dust distribution keeps a memory of its initial state because the coagulation timescale is long compared to the free-fall timescale. Conversely, the growth timescale is so short in the first Larson core that the initial conditions are quickly erased. In short, if the choice of initial dust distribution is of little importance for the fully coagulated state at high density, it matters nevertheless for the low density regions of the collapse (in the envelope).
 
 \subsection{Impact of the magnetic field strength}
 \label{sec:Bfield}
 \begin{figure}
\centering
     \includegraphics[width=
          0.4\textwidth]{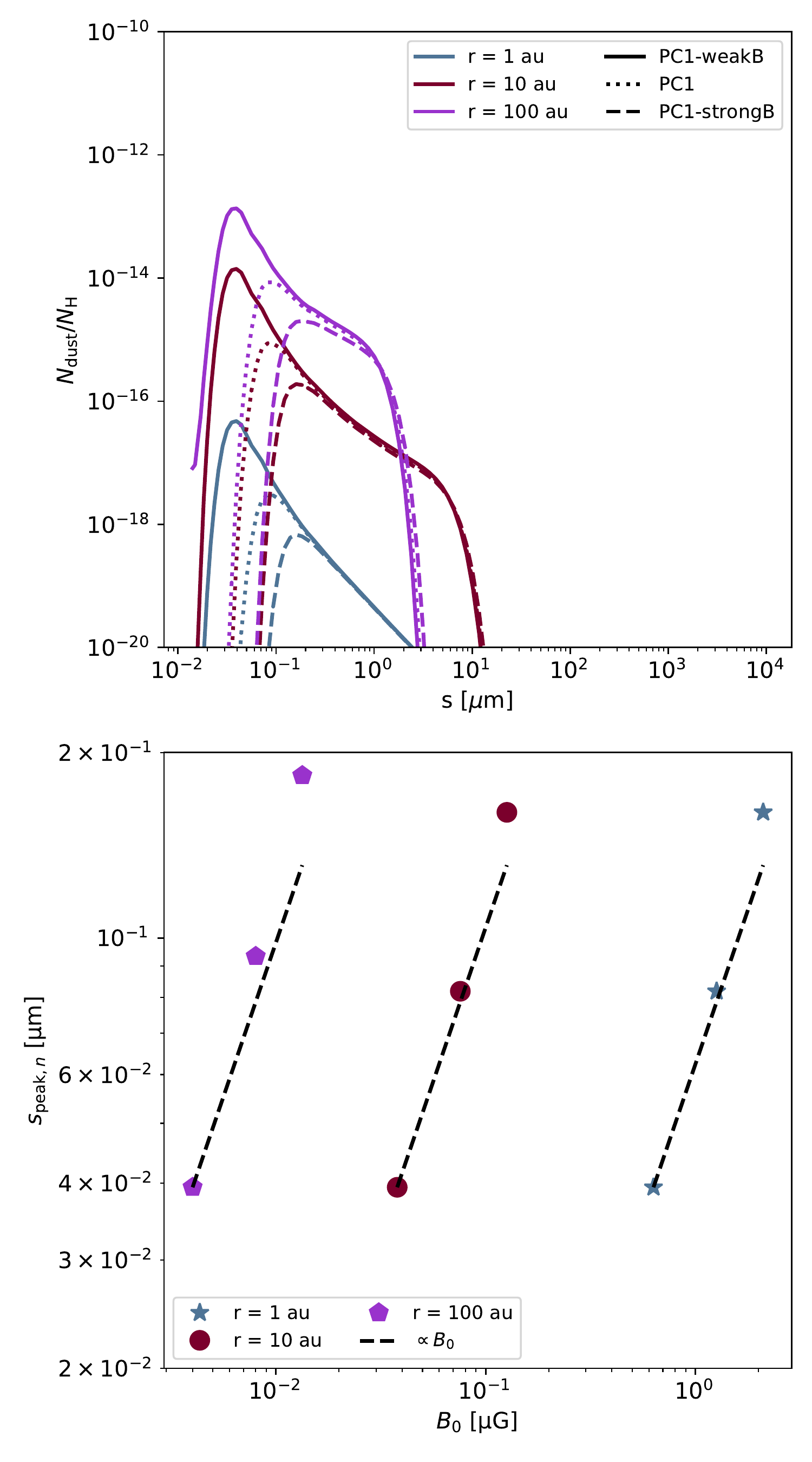}
     \caption{(Top) final dust number distribution for $\textsc{PC1-weakB}$ (plain line), $\textsc{PC1}$ (dotted line) and $\textsc{PC1-strongB}$ (dashed line) at r = 1 (blue), 10 (red) and 100 (purple) au. (Bottom) Position of the peak of the number distribution as a function of $B$ for the three positions.  }
                \label{fig:compB} 
\end{figure}
As we have seen above, the ambipolar diffusion is particularly efficient at removing the small grains in the case of $\textsc{PC1}$. However, this model was computed for a specific choice of magnetic field. We therefore computed two additional runs with a weaker ($\textsc{PC1-weakB}$, $B_0=15~\micro$G) and stronger ($\textsc{PC1-strongB}$, $B_0=50~\micro$G) outer magnetic field in order to see how it affects the small grain removal.
 
In the top panel of Fig.~\ref{fig:compB}, we show the dust number distribution of  $\textsc{PC1-weakB}$, $\textsc{PC1}$ and $\textsc{PC1-strongB}$ at various radii at the time of the first Larson core formation. We can clearly see the impact of the magnetic field strength on the abundance of small grains. As the magnetic field increases, the size under which grains are depleted also increases as ambipolar diffusion gets stronger. 
Still, the large grain tail of the distribution is almost unaffected by the change of magnetic field at all radius. This confirms again that the turbulent growth is the main responsible for large grain formation in our models. The value $s_{\mathrm{peak},n}$ of the peak of the number distribution (shown in the top panels with the vertical lines) is shown as a function of magnetic field strength at the three positions in the bottom panel of Fig.~\ref{fig:compB}. It corresponds to the size under which the small grains are depleted. As can be seen, at all radii, the small grains are increasingly more depleted with a stronger magnetic field. We see that $s_{\mathrm{peak},n}$ shifts from $\sim 4 \times 10^{-2}\,\micro\meter$ in $\textsc{PC1-weakB}$ up to  $\sim 2 \times 10^{-1}\,\micro\meter$ in $\textsc{PC1-strongB}$ that only has a stronger magnetic field by a factor of $\sim 3.33$. The dependency to the magnetic field seems to weaken with a decreasing radii (i.e. an increasing density). We indeed see that $s_{\mathrm{peak},n}$ has quasi linear dependency with $B$ that is more or less the same at the three radius (it is slightly shallower at low radius, i.e. at high density).

\subsection{Impact of the fragmentation}
\begin{figure*}
\centering
     \includegraphics[width=
          \textwidth]{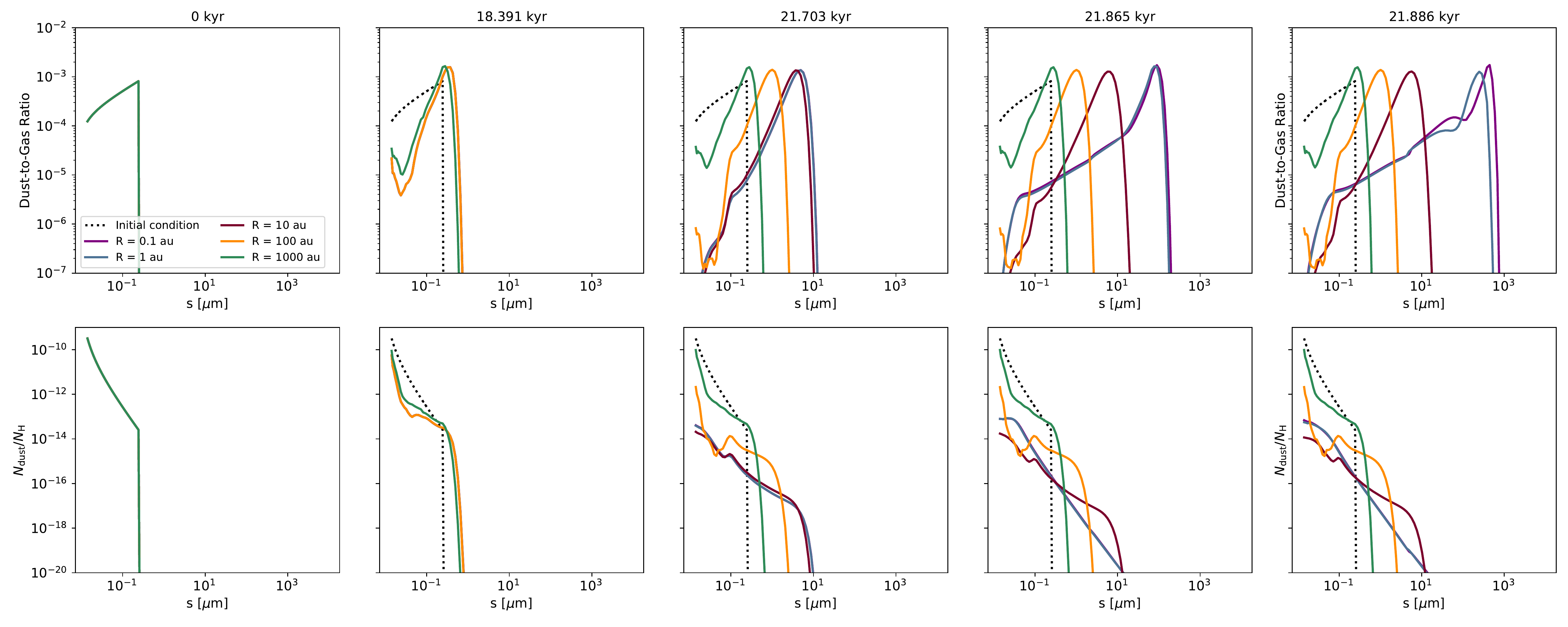}
     \caption{Same as Fig.\ref{fig:comp_vel} for run $\textsc{PC1-FRAG}$. The gas density evolution is not displayed as it is very similar to the one of $\textsc{PC1}$. }
            \label{fig:time_evol_frag} 

\end{figure*}
\label{sec:fragres}
As explained in Sect. \ref{sec:evolution}, collisions between grains become destructive above a certain energy threshold, leading to a redistribution of the target and projectile mass into fragments. Unfortunately, the outcome of this redistribution and the energy threshold for grain destruction are quite uncertain. In this work, and as a first step, we assumed that fragments are redistributed in a MRN-like power law with the recipes for fragmentation presented in Sect.~\ref{sec:context} and \ref{sec:method}.
We have computed two models with fragmentation. One considering the elastic properties of icy-grain, the result of which we do not present here because they are almost identical to $\textsc{PC1}$ with negligible fragmentation. And a second one, $\textsc{PC1-FRAG}$, computed with the elastic properties of bare-silicates. We show in Fig.~\ref{fig:time_evol_frag}, the same information as Fig.~\ref{fig:time_evol_fid}  but for the $\textsc{PC1-FRAG}$ model. The gas density evolution is not displayed because it is indistinguishable from the one of $\textsc{PC1}$.

As can be seen, $\textsc{PC1}$ and $\textsc{PC1-FRAG}$ are indeed quite different. Strong similarities can be observed at low densities, but even then the replenishment of the small grain population by the fragmentation is effective (which as we will see affects the resistivity profiles). At late times and high densities (low radii), we observe the well-known shape of the dust distribution obtained from an equilibrium between coagulation and fragmentation \citep{Birnstiel2011}. 

Let us focus on the description of the distribution at 0.1 au at the time t=21.866 kyr. Very distinctively, the dust accumulates at the fragmentation barrier around $100~\micro\meter$ and then, above it, the distribution sharply decreases. We note that this barrier corresponds to the most destructive collision, i.e. for a projectile and a target of the equal size. It thus corresponds to the commonly employed velocity threshold. The grains that are broken at the fragmentation barrier, and slightly below, are very effectively redistributed in a broad spectrum of sizes. This spectrum mainly ranges between 0.1 and $\sim 50~\micro\meter$. The slope of the distribution is very similar to the slope of the fragment that we imposed: the MRN power law index. It is worth mentioning that the distribution drops below $\sim 0.1~\micro\meter$, which is most likely due to the depletion of small grain by ambipolar diffusion that acts faster than the re-population by fragmentation for these sizes. In short, the distribution at fragmentation/coagulation equilibrium in the first Larson core can be decomposed into three regions:
\begin{itemize}
    \item For $50 ~\micro\meter < s < 100~\micro\meter $, the dust mass accumulates at the fragmentation barrier
    \item For $0.1 ~\micro\meter < s < 50~\micro\meter $, the dust is distributed according to the fragment power law distribution with a small bump near $50~\micro\meter$.
    \item For $s < 0.1 ~\micro\meter  $, the grains are depleted by ambipolar diffusion.
\end{itemize}

Another general observation should be made. We clearly see on the number distribution that the depletion of small grains is much less effective, at all densities, when fragmentation is included than in $\textsc{PC1}$. We will see later that this  impacts the resistivity profiles.

\section{Discussion}
\label{sec:discussion}
\subsection{Magnetic resitivities}
\label{sec:resi}
\begin{figure*}
\centering
     \includegraphics[width=
          0.9\textwidth]{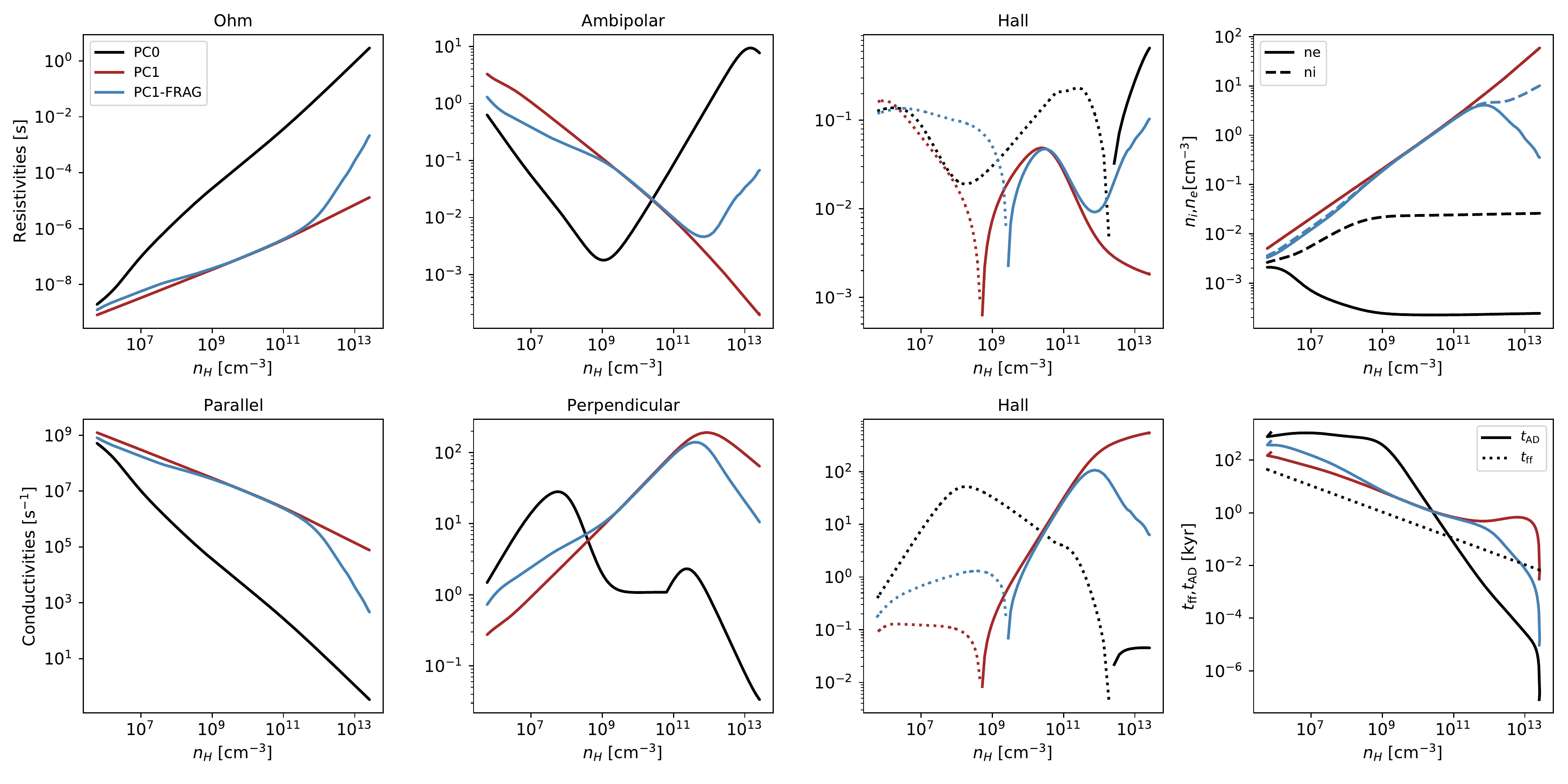}
     \caption{Top: Ohm (left), ambipolar (middle-left) and Hall (middle-right) resistivities and ion and electron densities (right) as a function of the density for $\textsc{PC0}$ (no grain growth, in black), $\textsc{PC1}$ (sticking only, in red) and $\textsc{PC1-FRAG}$ (sticking and fragmentation, in blue). Bottom: Parallel/Ohm (left), perpendicular (middle) and Hall (right) conductivities for the same models. In both cases, dashed lines correspond to the negative part of a quantities and plain lines are the positive parts.}
            \label{fig:resi_comp1} 

\end{figure*}

 One of the aims of this work is to constrain the magnetic resistivities during the protostellar collapse. We will focus on the comparison between $\textsc{PC0}$ (no grain growth), $\textsc{PC1}$ (coagulation only), $\textsc{PC1-FRAG}$ (coagulation and fragmentation).
 
 We show in Fig.~\ref{fig:resi_comp1} the profile of the resistivities, conductivities and the ion/electron densities as a function of the gas number density for the models at the end of the calculation. All these quantities are strongly affected by grain growth.  Let us first focus on the regions of low density, where the $\textsc{PC1}$ and  $\textsc{PC1-FRAG}$ models are  the most similar. In these regions, the Ohm resistivity is much lower with grain growth, whether fragmentation is included or not. This is because when grains coagulate their available surface area drops, which decreases their absorption of electrons. Consequently, more free electrons are present in the gas phase. As can be seen on the top-right panel, the electron fraction is indeed much higher in the case of $\textsc{PC1}$ and $\textsc{PC1-FRAG}$ than it is for $\textsc{PC0}$.

The effect of the grain growth on the ambipolar diffusion resistivity is also very strong at low densities, particularly when no fragmentation is considered. We see that the ambipolar diffusion is systematically stronger than in  $\textsc{PC0}$ up to densities of $\sim10^{10}-10^{11}~ \centi\meter^{-3}$. It even increases by almost 2 orders of magnitudes with respect to  $\textsc{PC0}$ around  $\sim10^{9} ~\centi\meter^{-3}$ for  $\textsc{PC1}$. This is particularly interesting because this is the typical density of the transition between the disk and the envelope in protostellar collapse calculations \citep[see for e.g.][]{Masson2016,Hennebelle2016,Hennebelle2020,Lebreuilly2021}. Such a high value of the ambipolar coefficient could promote the disk formation by strongly reducing the effect of the magnetic braking at intermediate scales \citep{Zhao2016,Marchand2020}. The increased impact of ambipolar diffusion during the collapse can also be seen in the bottom-right panel of  Fig.~\ref{fig:resi_comp1} that shows that the ambipolar timescale and the free-fall timescale are much more comparable up to densities of  $\sim10^{9}-\sim 10^{10}~ \centi\meter^{-3}$ when sticking is included (with and without fragmentation). We note that above  $\sim10^{11}~ \centi\meter^{-3}$ the opposite trend is observed for both $\textsc{PC1}$ and $\textsc{PC1-FRAG}$. They both have an ambipolar resistivity way below $\textsc{PC0}$, this could enhance the magnetic braking at high density. In short, if we summarise at all densities.  We could expect a low magnetic braking efficiency at low density associated with a strong regulation of the magnetic flux and an efficient braking at high density i.e., in the first core and the disk.
 
 If we now consider the impact of coagulation on the Hall resistivity (still at low densities), we see that the Hall effect is reduced by the coagulation and slightly increased by the early fragmentation. However, contrary to the Ohmic and ambipolar diffusion, the consequences of the Hall effect on the collapse will depend on the angle between the initial cloud angular momentum and magnetic field. It has indeed been  shown that the Hall effect could either enhance or reduce the effect of magnetic braking. In protostellar collapses, the Hall resistivity changes sign at disk-like densities, around $10^{12}-10^{13}$ cm$^{-3}$ \citep{Marchand2016,Wurster2016,Lee2021} for a MRN distribution (i.e., $\textsc{PC0}$ here). We see that this is also the case for the  $\textsc{PC1}$ and $\textsc{PC1-FRAG}$ models, but that the density at which it occurs is shifted. It occurs at $\sim 10^9 \centi\meter^{-3}$ for $\textsc{PC1}$ and at a slightly higher density for $\textsc{PC1-FRAG}$. This means that the inversion of the action of the Hall effect on the magnetic braking  \citep[strengthening or weakening it depending on the magnetic field direction][]{Marchand2018,Marchand2019,Zhao2021} would occur at a larger scale, i.e. in the envelope, when dust growth occurs.

We now investigate more in details the differences between $\textsc{PC1}$ and $\textsc{PC1-FRAG}$. The models start to strongly differ above densities of $\sim10^{11} \centi\meter^{-3}$ because collisions are not strong enough for significant fragmentation below this threshold. We insist again that there are still small differences between the two models even at low densities and that are quite significant for the Hall and ambipolar resistivity. This differences are due to a larger abundance of small grains in $\textsc{PC1-FRAG}$ due to early fragmentation.

We recall that the Ohmic resistivity is controlled by the presence of small grains, as the large surface area they provide causes the capture of many electrons from the gas phase. With only grain growth in $\textsc{PC1}$, the number of small grains strongly decreases with density, allowing electrons to flow more freely, and decreasing the Ohmic resistivity. For $\textsc{PC1-FRAG}$, once fragmentation starts to replenish more significantly the small grains populations (above $\sim10^{11} \centi\meter^{-3}$), the number of electrons drops (see right panel of figure \ref{fig:resi_comp1}), then the resistivity increases and eventually reaches a value closer to $\textsc{PC0}$. We note that, even then, it is still more than two orders of magnitude below that value.

A similar effect is observed in the ambipolar and Hall resistivity profiles, that are affected by the number of ions and the relative number of electrons and ions, respectively. In their case however, it is even significantly dropping for $\textsc{PC1}$. For $\textsc{PC1}$, the three resistivities are so low at high density that ideal MHD would probably be a good approximation here and magnetic braking could be very intense in the inner regions of protostellar collapse. Intense magnetic braking observed around some protostar, such as B335 for example \citep{Maury2018}, could be a clue that fragmentation is not that efficient at repopulating the small grain distribution at these stages (see Sect. \ref{sec:frag} for a discussion on fragmentation). Nevertheless, even if the resistivities are higher than $\textsc{PC1}$ for  above $\sim10^{11} \centi\meter^{-3}$ when fragmentation is included, they are still way below those of a MRN and might therefore be sufficient to explain the strength of the magnetic braking in B335.

It is important to mention that the resistivity profiles of these three models are in good agreement with the recent calculations of \cite{Kawasaki2022}. Despite some differences in the setup and method to solve the gas evolution, our models, $\textsc{PC1}$,  $\textsc{PC1-FRAG}$, and  $\textsc{PC0}$ are qualitatively similar to their models $\textsc{sil-coag}$,  $\textsc{sil-frag}$, and  $\textsc{MRN}$ and give similar distributions and resistivity profiles. We attribute the small changes between $\textsc{sil-frag}$ and $\textsc{PC1-FRAG}$ to our fragmentation recipe. We indeed use an energy threshold while they considered a unique velocity threshold which, as explained in Sect. \ref{sec:fragtheo} is equivalent, for a collision of equal-mass grains, to the choice of a lower energy threshold. The similarity of our results is particularly interesting since we solved the full hydrodynamical equations while they used a one zone model. We note that we do observe a variation of the resistivity profile over time that cannot be inferred in one-zone models. These variations of the resistivity profiles over time  are displayed for $\textsc{PC1}$ in the Appendix.~\ref{sec:appendix_resi}.

\subsection{Uncertainties in the modeling of fragmentation}
\label{sec:frag}

As shown in Sect.~\ref{sec:fragres} and \ref{sec:resi}, fragmentation could play a significant role during the collapse. That being said, the onset and outcome of fragmentation are however very unclear. First of all, the distribution of fragments is ill-constrained and might depend on the strength of the collision.  The outcome of fragmentation in laboratory experiments have be extensively reviewed by \citet{Guttler2010}. In their work they have shown that the power law index of the distribution of fragments could range between -2.1 and -1.2, which is significantly flatter than the MRN-like distribution we use here. \cite{Brauer2008} has however argued that a MRN power law index for the fragments was reproducing correctly the observed extinction curves in protoplanetary disks. With a shallower (resp. steeper) distribution of fragment than the MRN, we would expect less (resp. more) small grains than in $\textsc{PC1-FRAG}$ and therefore a stronger (resp. weaker) decoupling between the gas and the magnetic field. 

For simplicity, we assumed in our study that there exists no fragments smaller than $5~\nano\meter$, the minimal dust size of our grids. Changing this value for a lower one will have a significant impact on the number of dust grains produced by fragmentation since the smallest grains are the most abundant fragments in a MRN-like distribution. As a feedback, the resistivities will change significantly, since the abundance of small grains, that are also well coupled to the magnetic field, controls the numbers of free electron in the mixture. To which extent this effect could be compensated by the quick depletion of small grain population by ambipolar diffusion and brownian motion remains to be studied.

In addition, the fragmentation threshold depends on the elastic properties of the grains and the typical size of the monomers. We have found that fragmentation only takes place in the context of grains that have bare silicates properties. As we have also shown, the model that we computed with icy grains gave the exact same results as the $\textsc{PC1}$ model, i.e. the model without fragmentation.  $\textsc{PC1-FRAG}$ and $\textsc{PC1}$ (the results of the latter being identical to the icy-grains model), can thus be identified as the two extreme cases of fragmentation that may help us to provide lower and upper limits to the magnetic resistivities. Using the model with bare silicate grain is also physically motivated at high density. Fragmentation indeed mostly takes place at high density and high temperature (in the first core), where ice mantles might already be melting, thereby weakening the bonds between monomers in the aggregates. Replacing icy grains by bare silicates is therefore a simple way to model this drop in the resistance of aggregates to fragmentation. While our study is preliminary, more accurate future models should take into account the fact that the elastic properties of grain indeed depend on the condition at play in the cloud. 

As pointed out by \cite{Ormel2009}, the choice of the average monomer size is also key to the modeling of aggregate fragmentation. The fragmentation threshold of similar-sized grains scales as $s_{\mathrm{mono}}^{-5/6}$. With a lower (resp. larger) size, the grains would be more difficult (resp. easy) to break and fragmentation would therefore occur at higher (resp. lower) densities and the barrier would be shifted toward smaller (resp. larger) grain sizes. In fact, we have verified that considering the monomers twice smaller would totally suppress fragmentation even in the case of bare silicates. It is worth pointing out that the fragmentation threshold velocity obtained by \cite{Guttler2010} is about 1 $\meter ~\second^{-1}$. This is about an order of magnitude lower that what we find with monomers of $0.1~\micro\meter$ but is consistent with the almost invertly linear scaling of the threshold with the monomer size, these experiments indeed typically consider silicate grains $\sim 1~\micro\meter$. 

Finally, since we find that dust fragmentation is controlled by turbulence, it is important to point out the various uncertainties associated to the \cite{Ormel2007} model we use for the turbulent differential velocities. The expressions in  \cite{Ormel2007} model have been derived for a Kolmogorov cascade that is probably relevant during the protostellar collapse but not be in protoplanetary disks. In addition, it is considered that the energy is injected at the Jeans length, which simply might not be true for quiescent dense cores. We also assume $V_{\mathrm{g}} = \sqrt{3/2}c_{\rm{s}}$ as in previous studies, but this value should in principle depend on the turbulence level of the prestellar core, and as such might not be universal. It was also shown by \cite{Hennebelle2021} that the generation/amplification of the turbulence depends on the conditions at play in the dense core, i.e. its thermal support. In addition, MRI turbulence in protoplanetary disks might lead to larger differential velocities than the ones of a Kolmogorov cascade \citep{Gong2020,Gong2021}. On top of that, the \cite{Ormel2007} model does not consider that the turbulent cascade can be affected by the dust back-reaction onto the gas. This could be wrong at small scales (where coagulation happens) and should therefore be investigated in future works. Last but not least, the intermittency of turbulence might be extremely important for the population of small grains, rare events of high collision velocity could lead to a replenishment of the small grain populations even if the average kinetic energy is below the fragmentation threshold. The assumption on the turbulent properties would not only impact the dust coagulation timescale but also the position of the fragmentation barrier.

Understanding the conditions and consequences of fragmentation is key in the context of protostar, protoplanetary disk and planet formation. 
A shifting of the fragmentation barrier would not only have consequences on the magnetic resistivities, but also on the onset of planet formation since larger grains would be more likely to trigger instabilities such as the streaming instability in disks \citep[see][and subsequent works on the streaming instability]{YoudinGoodman2005,JohansenYoudin2007}. Constraining grain fragmentation in cores, as well as the turbulent velocity, thus appears urgent and of great importance. These issues require fully dedicated works and are, as such, way beyond the scope of this paper. They will however be carefully investigated in forthcoming focused studies.

\subsection{Magnetic flux regulation by the removal of small grains}
\begin{figure}
\centering
     \includegraphics[width=
          0.35\textwidth]{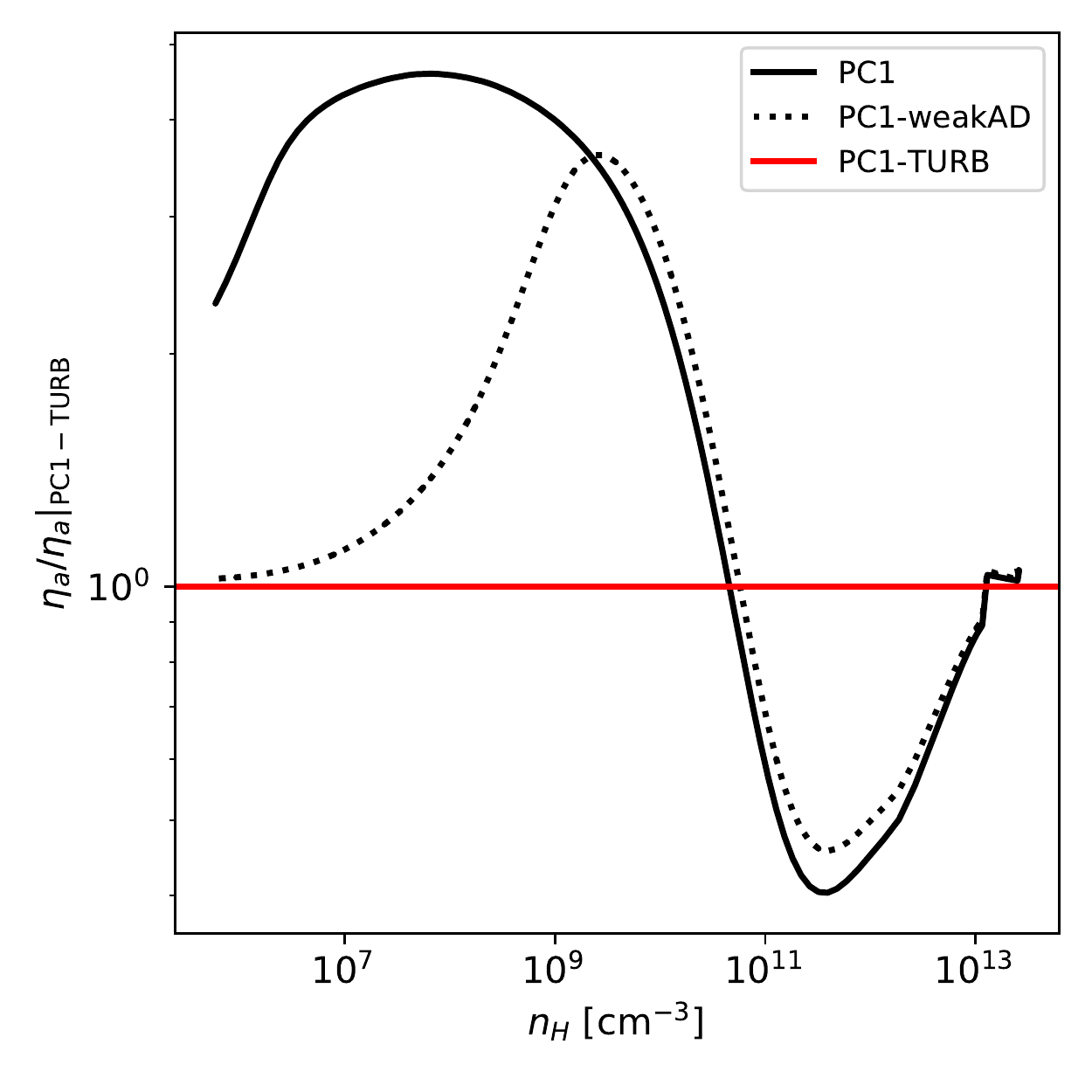}
     \caption{$\Theta\equiv {\eta_{AD}}/{\eta_{AD}}|_{\mathrm{PC1-TURB}}$ as a function of the density for $\textsc{PC1-weakAD}$ (plain line) and $\textsc{PC1}$. }
            \label{fig:theta_res} 

\end{figure}
 As proposed by \cite{Guillet2020}, the magnetic flux could be regulated by the abundances of small grains in prestellar cores if a stronger (resp. weaker) diffusion of the field lines occurs when the small grain population is enhanced (resp. depleted). We have investigated the impact of the values of the magnetic field in Sect.~\ref{sec:Bfield} and our finding are consistent  with this tentative scenario.

However to be more conclusive, we need to verify that depleting the small grain population does indeed increase the ambipolar resistivity. In order to verify that, we computed an additional model $\textsc{PC1-weakAD}$ with $\delta=0.1$, i.e. with a lower ambipolar velocity than $\textsc{PC1}$. This results in a reduced depletion of the small grain. We now examine the impact of the change of small grain population on the ambipolar resistivity by displaying the ratio $\Theta \equiv \eta_{\mathrm{AD}}/{\eta_{AD}|_{\mathrm{PC1-TURB}}}$ for $\textsc{PC1}$ and $\textsc{PC1-weakAD}$ where $\eta_{AD}|_{\mathrm{PC1-TURB}}$ is computed with the $\textsc{PC1-TURB}$ model. This quantity is shown for the two models as a function of density in Fig.~\ref{fig:theta_res}. We remind that grain grow significantly in $\textsc{PC1-TURB}$ but that the small grain are not depleted at all, hence we can attribute the variation of ambipolar resistivity solely to the changes in small grain populations.

As can be seen, at density lower than $\sim 10^9~ \centi\meter^{-3}$, removing the small grains provokes a significant increase of the ambipolar resistivity and we thus expect the magnetic field to be diffused more when the small grains are depleted. At higher densities (in the first core), the removal of small grains has the opposite effect, but as explained in Sect.\ref{sec:resi}, the resistivity is probably so low there that ideal MHD could be a good approximation.

If we summarise what happens for density lower than $\sim 10^9~\centi\meter^{-3}$, increasing the magnetic fields does enhance the small grain removal (see Sect.~\ref{sec:Bfield}) and removing the small grains does increase the resistivity (see Fig.~\ref{fig:theta_res}). The two conditions for the regulation of the magnetic flux during the protostellar collapse by small grains are typically met up to first core densities.  

\subsection{Toward multidimensional calculations}

Real protostellar collapses are not spherically symmetrical, dense cores are indeed rotating and momentum conservation tends to flatten structures and leads, in this case, to the formation of a protoplanetary disk aroung the protostar. This disk is of particular interest since it is where the planets are expected to form. Only a few studies have investigated dust growth in 3D (or even in 2D). In particular, \cite{Bate2022} recently investigated the effect of rotation with the first 3D hydrodynamical calculations of protostellar collapse that included a solving of the Smoluchowski equation. The comparison of our model to theirs is not straightforward because of the different setups and because we do not integrate the collapse beyond densities of $10^{-10}~\gcc$, but it is useful to recall their main findings: grains grow to a few microns prior to the first core formation and the growth is then accelerated. This is consistent with our findings, although the main mechanism to grow large grains is brownian motions in their work, while it is turbulence in our case (see Sect~\ref{sec:impact_source_velocity} for an explanation of this difference). \cite{Bate2022} have also shown that rotation could enhance the effect of grain growth in the spirals of disks. Similar findings have been found before in the context of 2D calculations \cite{Vorobyov019,Vorobyov2019,Elbakyan2020}. Unfortunately, dust growth simulations are still very expensive in 2D/3D because they either require a large number of dust species to converge or to use very advanced numerical methods \citep{Lombart2021,Lombart2022}.

An interesting approach  proposed by \cite{Stepinski1996} has recently been employed in the context of the protostellar collapse by \cite{Tsukamoto2021}. It consists in following a single dust species of evolving size, advected as a passive scalar in the model. Although it does not allow to fully evolve the dust distribution, it is already a very interesting approach to follow the peak of the dust mass distribution and therefore to estimate the dust mass content of protoplanetary disks.  \cite{Silsbee2020} have in fact shown that the peak of the distribution was quite well (although not perfectly) reproduced by the \cite{Stepinski1996} model. We see in our models (when fragmentation can be neglected) that the distribution resembles a power law ranging between the minimal grain size, controlled by ambipolar diffusion and the maximal grain size, controlled by turbulence. Hence monodisperse calculations could be used to extrapolate, in an approximated way, the complete dust distribution. The case when fragmentation occurs is more complicated, but we recall that the shape of a distribution at equilibrium between fragmentation and coagulation has also been derived \citep{Birnstiel2009}. 

Finally, it was shown by \cite{Marchand2021} that, as long as fragmentation can be neglected and if the dependency of the growth kernel on the gas and dust can be separated, then the coagulation becomes a 1D process and the evolution of the dust distribution can be parameterized simply as a function of a single variable $\chi$. In the context of pure turbulence, they have shown $d\chi = n_{\mathrm{H}}^{3/4} T^{-1/4} dt$. Using pre-computed coagulation tables parameterized as a function of $\chi$ thus simply allows to evaluate the dust growth during the collapse as long as the aforementioned approximations are valid.
In spite of these approximations, this method is a very inexpensive solution to include the dust growth and is a step further toward fully consistent models since it is compatible with an 'on-the-fly' estimate of the magnetic resistivity in 3D calculations.  

\section{Conclusion}
\label{sec:conclusion}

In this work, we investigated the coagulation and fragmentation of dust grains during the protostellar collapse using our newly developed {\ttfamily shark} code. We recall here our main findings:
\begin{itemize}
    \item The coagulation of dust grains is not negligible during the protostellar collapse and is accelerated at high density. Over a single free-fall timescale, grains grow up to a few tens of $\micro\meter$ in the protostellar envelope and beyond $100~\micro\meter$ in the initial stages of first Larson core.
    \item Under our assumptions, for grains with bare silicates elastic properties, we find that the fragmentation of grains is extremely important at high densities and is not completely negligible at low densities. For icy-grains, fragmentation is completely negligible during the collapse.

        \item The evolution of the dust size distribution through the competition of coagulation and fragmentation strongly impacts the value of all the magnetic resistivities and the abundances of ions and electrons.
 
\item Turbulence is found to be the main mechanism to form large grains, should grains be in the intermediate coupling regime, an assumption that depends on the scaling of the Reynolds number with the local cloud conditions.
\item Both the ambipolar diffusion and brownian motions are capable of removing the small grains from the distribution as standalone processes. Unlike ambipolar diffusion, brownian motions are inefficient in that matter when turbulence is included.  
\item We find the hydrodynamical drift due to the imperfect coupling between the gas and the dust to be negligible for growing grains and leading to only very small variation of dust-to-gas ratio in the accretion shock of the first Larson core.
    \item The choice of the initial distribution has little impact on the final coagulated state at high densities (in the first Larson core) but is important in the envelope where the grain growth timescale is longer.
    \item We have found that increasing the magnetic field strength enhances the removal of small grains by increasing their ambipolar drift. Additionally, removing the small grains increases the ambipolar diffusion resistivity and the ambipolar drift at densities lower than those of the first Larson core. This is new evidence that the magnetic flux could be regulated by the abundance of small grains during the protostellar collapse, as suggested by \cite{Guillet2020}.
         \item We have made a particular effort to understand the uncertainties in the modeling of fragmentation. This highlighted the non-linear dependence of the results with the choice yet ill-constrained physical quantities such as the typical monomer size, the elastic properties of the grains or even the differential turbulent velocity.
\end{itemize}

  \section*{acknowledgements}
We thank the referee for providing very useful comments that helped us to improve our manuscript. U. Lebreuilly and V. Vallucci-Goy acknowledge financial support from the European Research Council (ERC) via the ERC Synergy Grant ECOGAL (grant 855130). M. Lombart acknowledges the financial support from the Ministry of Science and Technology, Taiwan (MOST 110-2636-M-003-001). We thank P. Hennebelle for the stimulating discussions.

\section*{Data Availability}
The data underlying this article will be shared on reasonable request to the corresponding author. Upon publication the models will be shared in the Galactica database (\url{http://www.galactica-simulations.eu/db/}).
\bibliographystyle{mnras}
\bibliography{ref}

\appendix
\section{Modeling the initial distribution}
\label{sec:appendix_distri}
\subsection{MRN}
\label{sec:appendix_MRN}

We initialise the dust-to-gas ratio $\epsilon_j$ of each bin $j$ of sizes between $s_{\mathrm{min, init}}  \equiv 5 ~\nano\meter$ and $s_{\mathrm{max, init}} \equiv 250 \nano\meter$ according to a MRN distribution \citep{MRN}  such as
\begin{equation}
\epsilon_j \equiv \epsilon_0 \frac{s_{+,i}^{\lambda+4}-s_{-,i}^{\lambda+4}}{s_{\mathrm{max, init}}^{\lambda+4}-s_{\mathrm{min, init}}^{\lambda+4}},
\end{equation}
where $\lambda$ ($\lambda=-3.5$ for a standard MRN)  is the power law index of the MRN distribution. Since we allow grains to grow, the grid range $[s_{\mathrm{min}},s_{\mathrm{max}}]$ is larger than the range $[s_{\mathrm{min, init}},s_{\mathrm{max, init}}]$, the dust bins that are outside of this range  are thus seeded with a very small dust-gas-ratio of $10^{-15}$ that is negligible but not zero to avoid numerical instabilities.

\subsection{Log-Normal}
\label{sec:appendix_logn}
Both theory and observations suggest that a log-normal distribution might be more accurate than a power law for the dense regions of the ISM. We therefore also investigate the evolution of such distributions. Considering a mean grain size $s_{\mathrm{mean}}$ and a standard deviation $\sigma$ than we simply have 
\begin{equation}
\epsilon_j \equiv \epsilon_0 \frac{\rm{erf} \left(\frac{1}{\sqrt{2 \sigma}} \rm{log} \left(\frac{s_{+,i}}{s_{\rm{mean}}}\right) \right) -\rm{erf} \left(\frac{1}{\sqrt{2 \sigma}} \rm{log}\left( \frac{s_{-,i}}{s_{\rm{mean}}}\right) \right)}{\rm{erf} \left(\frac{1}{\sqrt{2 \sigma}} \rm{log} \left(\frac{s_{+,i}}{s_{\rm{max}}} \right)\right) - \rm{erf} \left(\frac{1}{\sqrt{2 \sigma}} \rm{log}\left( \frac{s_{min}}{s_{\rm{mean}}}\right) \right)},
\end{equation}
we note that contrary to the MRN distribution, the Log-Normal one extends to the whole initial range of dust sizes. 
\subsection{Pre-grown}
\label{sec:appendix_preg}
Another way to consider that dust growth starts before the protostellar collapse is to model the growth (with all the velocity sources except the hydrodynamical drift\footnote{it cannot be included in such simulation as it requires to solve the hydrodynamical evolution of the cloud}) of a distribution, here the MRN, in typical dense core conditions for a given amount of time. Our so-called GROWN distribution are extracted from single-cell static {\ttfamily shark} simulations that assume an initial standard MRN distribution. The final distribution then depends on the conditions at play in the cell, i.e. the choice of density ${\nh}_{0}$ (the temperature being computed with the barotropic equation of states presented in Sect.\ref{sec:setup}), and time during which the distribution is evolved $t_0$. In this work, we explored one GROWN ditribution presented in Table~\ref{tab:dust-distri}.

\section{Testing {\ttfamily shark}}

In this section, we present several tests that we performed to validate the implementation of the different elements of {\ttfamily shark}.

\subsection{Gas hydrodynamics - comparison with ramses}

\begin{figure}
\centering
     \includegraphics[width=
          0.45\textwidth]{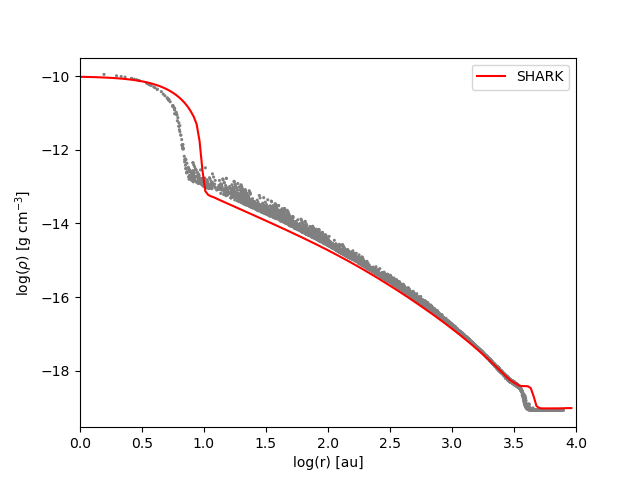}
     \caption{{\ttfamily shark} (red) vs {\ttfamily ramses} (grey). Density as a function of the radius when the maximum density reaches $10^{-10}\gcc$. }
            \label{fig:sharkvsramses} 
\end{figure}

To test our implementation of the collapse physics in {\ttfamily shark}, we compare the results we obtained with a similar collapse calculation with the {\ttfamily ramses} code \citep{Teyssier2002}. For this comparison, we used the same barotropic equation of state as the one implemented in  {\ttfamily ramses} which gives
\begin{equation}
T=T_{0} \left(1+\left(\frac{\rho}{10^{-13}\gcc}\right)^{\gamma-1}\right).
\end{equation}
In our {\ttfamily ramses} calculation, computed in 3D, we used an adaptive-mesh refinement grid of base resolution $64^{3}$ and a range of $8$ levels of refinement imposing at least 10 points per Jeans length.

As can be seen in Fig.~\ref{fig:sharkvsramses}, that shows the density profiles obtained with the two codes when the peak density reaches $10^{-10}~\gcc$, the results are in good agreement despite a very different choice of grid and geometry. There is $99\%$ agreement between the free-fall timescales of the two models. We also verified that mass is conserved to machine precision in {\ttfamily shark} when no inflow and outflow are allowed.

\subsection{Charging - comparison with ishinisan}

\begin{figure}
\centering
     \includegraphics[width=
          0.45\textwidth]{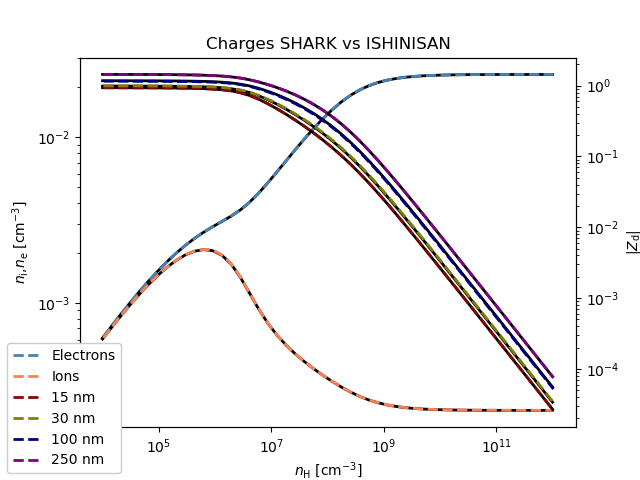}
     \caption{Charging. Comparison of the electron, ion density and grain charges obtained with {\ttfamily shark} (colored dashed lines) and  {\ttfamily ishinisan} \citep[plain black lines][]{Marchand2021}.}
            \label{fig:sharkvsishinisan} 

\end{figure}
We now present a test of the charge solver against the one of {\ttfamily ishinisan} \citep{Marchand2021}. Both codes use the  method of \cite{Marchand2021} to compute the ionisation. We note that, as pointed out by \cite{Tsukamoto2021}, the method does not converge when the dust charge density goes to zero, but we have found a simple solution for that issue. We simply consider that the fraction $\frac{n_e}{n_i}$ is always smaller that $\epsilon_{max}=1-\delta$, with $\delta \ll 1$. When $\frac{n_e}{n_i}=\epsilon_{max}$, i.e., when the dust charge density becomes negligible, the recombination of ions and electrons onto dust grains are negligible and 
\begin{equation}
n_i = \sqrt{\frac{\epsilon_{max} \zeta n_{\mathrm{H}}}{\left<\sigma_{v,\mathrm{ie}}\right>}},
\end{equation}
where $\left<\sigma_{v,\mathrm{ie}}\right>$ is the collision rate between ions and electrons, determined as in \cite{Marchand2021}.

To compare the codes we perform a charge computation for a MRN distribution (with 50 bins) and $1\%$ of dust in a number density ranges between $10^{4}~\centi\meter^{-3}$ and $10^{12}~\centi\meter^{-3}$ at a temperature of 10 K. We use the same ionisation rate, ion mass and sticking efficiency of electrons onto grains as in the rest of the paper. In figure~\ref{fig:sharkvsishinisan}, we show the electron and ion density as a function of the number density, as well as the charge of three dust species obtained with {\ttfamily shark} (dashed lines). The reference by {\ttfamily ishinisan} is displayed in plain black lines. As can be seen, the agreement between the two codes is perfect.

\subsection{Dust hydrodynamics - Dust settling}

\begin{figure}
\centering
     \includegraphics[width=
          0.45\textwidth]{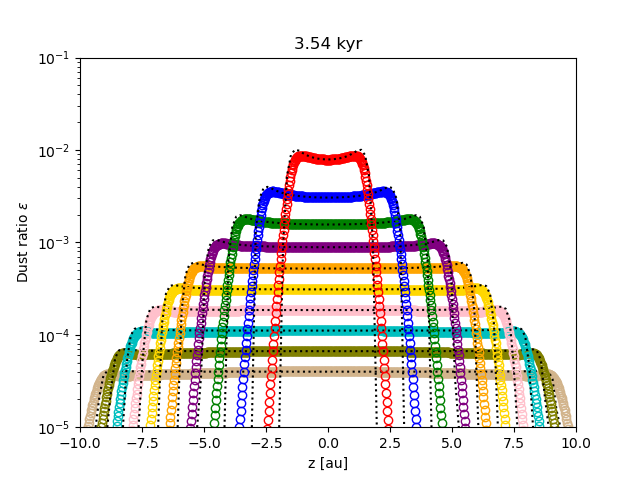}
     \caption{Dust settling. Dust ratio of the ten different species after 10 orbits obtained with {\ttfamily shark} (color circles) compared with the semi-analytical solution (black dotted lines).}
            \label{fig:stream} 

\end{figure}

We tested the implementation of dust dynamics by investigating the settling of dust grains in a stratified envelope of protoplanetary disk \citep{PriceLaibe2015,Hutchison2018,Lebreuilly2019}. We used the exact same setup and solutions as the ones described by \cite{Lebreuilly2019}. We performed this test with a uniform grid with a resolution of $0.01~$au. We employ zero-gradient with no inflow boundaries, to limit their impact we modelled the disk across 6 scale heights above and below the mid-plane. Figure~\ref{fig:stream} shows the dust ratio profile of the ten dust species that we considered as a function of height after 10 orbits against a semi-analytical solution (the same as in \cite{Hutchison2018,Lebreuilly2019}). The increasingly settled profiles correspond to grains of increasing sizes (from micrometer to centimeter, see Table 1 of \cite{Lebreuilly2019}). The choice of color coding is also exactly the same, we thus refer the reader to these studies for more details on this test. As can be seen, the analytical solution is very well reproduced by our {\ttfamily shark} simulation. We emphasize that, contrary to \cite{Lebreuilly2019}, we here solve the multifluid equations for the gas and dust without any terminal velocity approximation. Note that this test is not only a test for the dust, but also for the gas that stays at equilibrium here under the balance of thermal pressure and gravity.

\subsection{Dust growth}
\label{sec:appendix_growth_ana}
\begin{figure*}
\centering

\centering
  \begin{subfigure}[b]{0.33\textwidth}
  \centering
 \includegraphics[width=
          \textwidth]{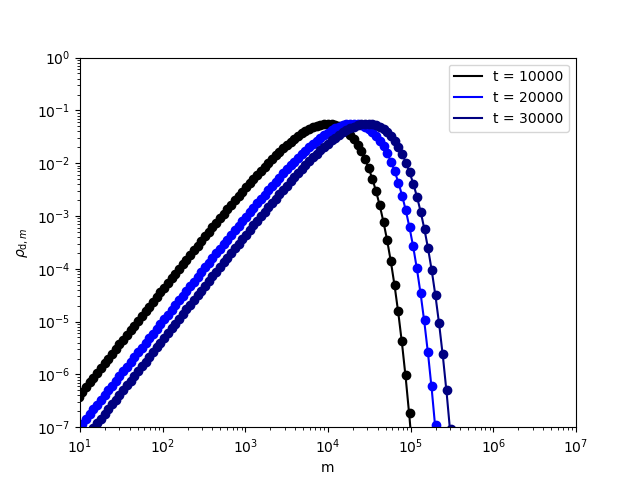}
      \caption{Constant kernel}
       \label{fig:sharkconstant} 
     \end{subfigure}
     \begin{subfigure}[b]{0.33\textwidth}
  \centering
 \includegraphics[width=
          \textwidth]{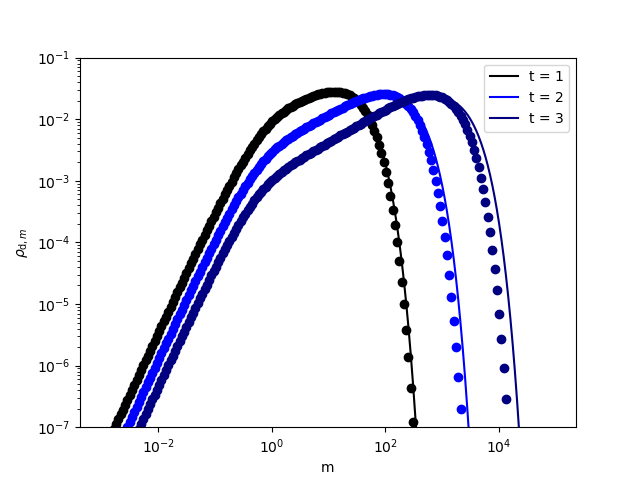}
      \caption{Additive kernel}
       \label{fig:sharkadd} 
     \end{subfigure}  
         \begin{subfigure}[b]{0.33\textwidth}
  \centering
 \includegraphics[width=
          \textwidth]{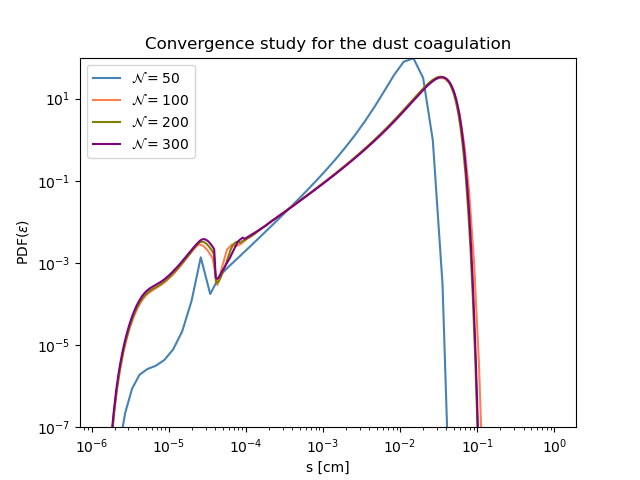}
      \caption{Physical kernel}
       \label{fig:sharkconv} 

     \end{subfigure}
       \caption{[Left and middle] Dust distribution  as a function of the grain mass at various times for the constant (left) and additive (middle) kernels. Result from {\ttfamily shark} (circles) compared with the analytical solution. [Right] Convergence study of the dust normalised probability distribution function in the presence of all the source of relative velocity (except the hydrodynamical drift). }
          \label{fig:analyticalsol}

\end{figure*}
 Solutions of the Smoluchowski equation can be derived in the context of the constant $K_{i,j}=1$ \citep{Scott1968,Silk1979} and additive $K_{i,j}=m_i + m_j$ \citep{Scott1968,Safronov1972}  kernels. The setup and the analytical solutions are the same as in 
\cite{Lombart2021}. We performed both tests with $\mathcal{N}=100$. For the constant (additive) kernel, we considered a mass range for the dust between $0.1$ and $10^7$ ($10^5$). The results are shown in the left and middle panels of Fig.~\ref{fig:analyticalsol}.  In both tests, the analytical solution is well reproduced in the considered range of time. We note a slight disagreement at the high mass tail for the additive kernel. This is a well-known and recurrent problem of most Smoluchowski solvers, that can only be fixed through high order methods \citep{Lombart2021,Lombart2022}.
 
We cannot compare the solution of the Smoluchowski equation in the context of realistic kernels, however it is possible to test for the convergence of the algorithm. We therefore performed single cell simulations of growth with all the sources of growth (hydrodynamical drift excluded) with $50$, $100$, $200$ and $300$ bins and the same grid as in the protostellar collapse runs. The method is the same as the one used to compute the GROWN distributions, we considered here $N_{\mathrm{H},0}=10^{10} \centi\meter^{-3}$ and integrate the simulations for 5000 years. As can be seen, we obtain a satisfying convergence of the dust distribution for $\mathcal{N}>100$. We note the position of a bump at $10^{-5}~\centi\meter$ that is due to the change of regime of turbulence seen by the grain with an increasing size (from class I to II).

\section{Time variation of the resistivities}
\label{sec:appendix_resi}
\begin{figure*}
\centering
     \includegraphics[width=
          \textwidth]{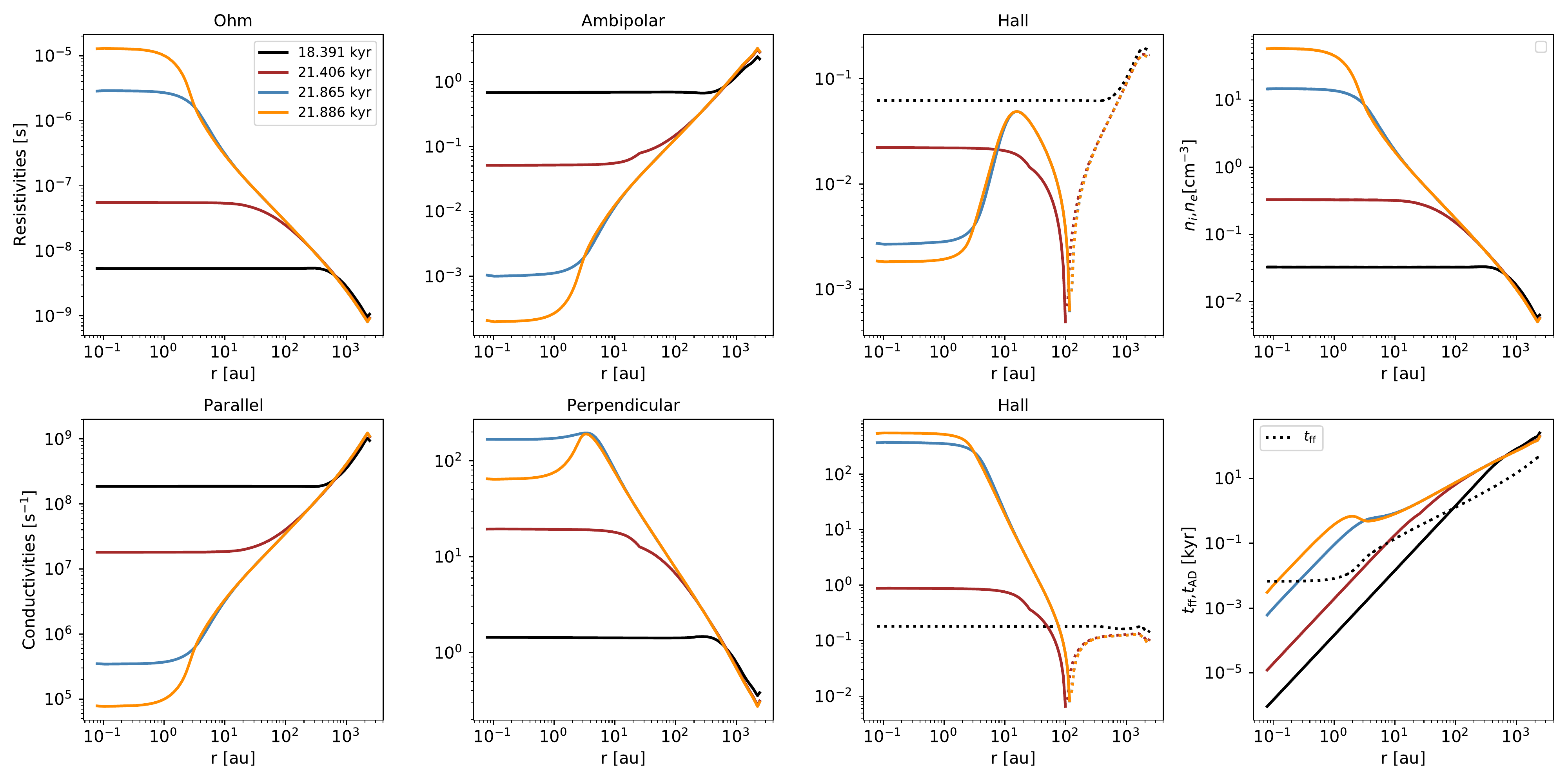}
     \caption{Top: Ohm (left), ambipolar (middle-left) and Hall (middle-right) resistivities and ion and electron densities (right) as a function of the radius for  $\textsc{PC1}$. Bottom: Parallel/Ohm (left), perpendicular (middle) and Hall (right) conductivities for the same models. In both cases, dashed lines correspond to the negative part of a quantities and plain lines are the positive parts.}
            \label{fig:resi_profile_time} 

\end{figure*}

We display here the variation of the resistivities, ion and electron densities, conductivities and ambipolar and free-fall timescale as a function of the radius at various time for the model $\textsc{PC1}$. We see that all these quantities greatly vary over time. 
\bsp	
\label{lastpage}
\end{document}